\RequirePackage{fix-cm}
\documentclass[twocolumn,epjc3]{svjour3} 
\smartqed  
\RequirePackage{amssymb}
\RequirePackage{amsmath}
\RequirePackage{array} 
\RequirePackage{epsfig}
\RequirePackage{graphics}
\RequirePackage{colortbl}
\RequirePackage{hhline}
\RequirePackage{multirow}
\begin{document}


\title{Central exclusive diffractive production of two-pion continuum at hadron colliders
}


\author{R.A.~Ryutin \thanksref{e1,addr1}
}

\thankstext{e1}{e-mail: Roman.Rioutine@cern.ch}


\institute{{\small Institute for High Energy Physics, NRC ``Kurchatov Institute'', Protvino {\it 142 281}, Russia} \label{addr1}
}

\date{Received: date / Accepted: date}

\maketitle

\begin{abstract}
\noindent Calculations of central exclusive diffractive di-pion continuum production are presented in the Reg\-ge--eiko\-nal
approach. Data from ISR, STAR, CDF and CMS were analyzed
and compared with theoretical description. We also
consider theoretical predictions for LHC, possible nuances and problems of calculations and pros\-pects of investigations at present and future hadron colliders.
\end{abstract} 
%
%

\section*{Introduction}

In previous papers~\cite{myCEDP1},\cite{myCEDP2} the general
properties and calculations of the Central Exclusive Diffractive
Production (CEDP) were considered. It was shown, especially
in~\cite{myCEDP2}, that diffractive patterns (differential cross-sections) of CEDP play a significant role in model ve\-ri\-fi\-ca\-tion.

Here we partially continue the subject of~\cite{myCEDP2} and investigate
in detail the process of low mass CEDP (LM CEDP) with
production of two pions. This process is one of the ``standard candles'' for
LM CEDP. Why do we need exact calculations and predictions for this process?
\begin{itemize}
\item Di-pion LM CEDP is the basic background process for
CEDP of resonances (like $f_2$ or $f_0$), since one of the basic hadronic decay modes for these resonances is the two pion one.
\item We can use LM CEDP to fix the procedure of calculations of ``rescattering'' (unitarity) corrections. In the case of di-pion LM CEDP there
are two kinds of corrections, in the proton--proton and the pion--proton
subamplitudes. They will be considered in the present work.
    \item The pion is the most fundamental particle in the strong interactions, and LM CEDP gives us a powerful tool to thoroughly investigate its properties, especially to investigate the form factor and scattering amplitudes for the off-shell pion.
    \item LM CEDP has rather large cross-sections. It is very important for an exclusive process, since in the special low luminosity runs (of the LHC) we need more time to get enough statistics.
    \item As was proposed in~\cite{mySD}, it is possible to extract some
    Reggeon--hadron cross-sections. In the case of single and double dissociation
    it was the Pomeron--proton one. Here, in the LM CEDP of the di-pion we can
    ana\-ly\-ze properties of the Pomeron--Pomeron to pion--pion exclusive
    cross-section, and also check again the predictions of the covariant reggeization method~\cite{mySD}.
    \item Diffractive patterns of this process are very sensitive to different
    approaches (subamplitudes, form factors, unitarization, reggeization procedure), especially differential cross-sections in $t$ and $\phi_{pp}$ (azi\-mu\-thal angle between final protons), and also the $M_{\pi\pi}$ dependence. That is why this process is used to verify different models of diffraction.
    \item All the above items are additional advantages provided by the LM CEDP of two pions, which has the usual properties of CEDP: a clear  signature  with  two final protons and two  large  rapidity  gaps\linebreak  (LRG)~\cite{LRG1},\cite{LRG2}  and the  possibility  to  use  the  “missing  mass  method”~\cite{MMM}.
\end{itemize}

\begin{figure}[hbt!]
\begin{center}
\includegraphics[width=0.37\textwidth]{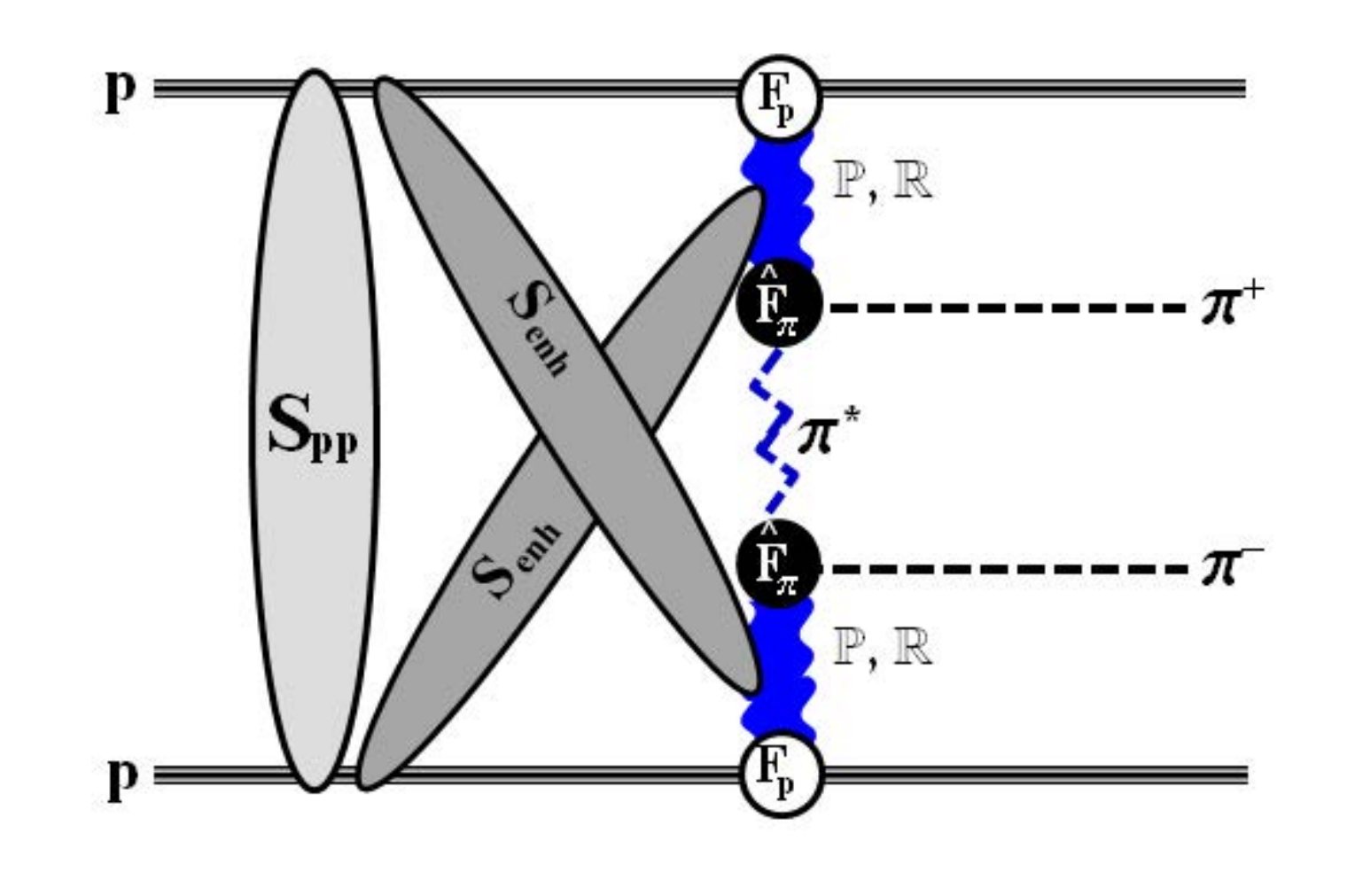}
\caption{\label{fig1:KMR} Amplitude of the process of double pion LM CEDP $p+p\to p+\pi^++\pi^-+p$ in the KMR approach~\cite{CEDPw1}-\cite{CEDPw3}. The central part of the diagram is the Born amplitude (with Pomeron and two reggeons). A reggeized
off-shell pion propagator is shown as a dashed zigzag line. Proton--proton rescattering is depicted as the $S_{pp}$-blob, and ``enhanced'' corrections
are also shown as the $S_{enh}$-blob. The off-shell pion form factor is presented as a black circle.}
\end{center}
\end{figure}

Processes of the LM CEDP were calculated in some other work~\cite{CEDPw1}-\cite{CEDPw6} devoted to the most popular models
for the LM CEDP of di-mesons. All authors have considered
nonperturbative approach in reggeon--reggeon collision subprocess. For example,
in the Dur\-ham\linebreak model~\cite{CEDPw1}-\cite{CEDPw3} (see Fig.~\ref{fig1:KMR}), they
take the Born Regge term for the amplitude of the process $p+p\to p+\pi^++\pi^-+p$ with reggeized propagator of the off-shell pion, then
they take into account unitarity (rescattering) corrections
(in the initial proton--proton state and also in the so-called ``enhanced'' one). The authors of~\cite{CEDPw1}-\cite{CEDPw3}
point out, however, that ``enhanced'' corrections are negligible due to the small triple Pomeron
vertex. In addition, the possibility to include reggeization
of the virtual pion propagator is not obvious, since the effect of this 
is expected to be small and, moreover, it is not even clear
that we are in the relevant kinematic region ($|\hat{t}|\ll \hat{s}=M_{\pi\pi}^2 $) to include such corrections for central
production. This is therefore not included by default into their 
calculations (see~\cite{CEDPw3}, for example). For example, the authors of~\cite{CEDPw1}-\cite{CEDPw3} 
use the replacement
\begin{equation}
\frac{1}{\hat{t}-m_{\pi}^2}\to \frac{\mathrm{e}^{\alpha_{\pi}(\hat{t})|\Delta Y|}}{\hat{t}-m_{\pi}^2},
\label{eq:KMRpionprop}
\end{equation}
which gives the correct ``reggeized'' behavior in the re\-le\-vant kinematical region, and 
the usual ``bare'' pion propagator behavior for a small difference between rapidities 
of pions. The authors of~\cite{CEDPw4}-\cite{CEDPw6} 
use  a phenomenological expression for the virtual pion propagator  (see~(\ref{eq:PionPropagatorR}) for notations, and 
also (3.25) and (3.26) of~\cite{CEDPw5a}) like
\begin{eqnarray}
&& \frac{1}{\hat{t}-m_{\pi}^2} F(\Delta Y) + (1-F(\Delta Y)) {\cal P}_{\pi}(\hat{s},\hat{t}),\nonumber\\
&& F(\Delta Y)=\mathrm{e}^{-c_y \Delta Y},\, \Delta Y=y_{\pi^+}-y_{\pi^-}, 
\label{eq:LIBpionprop}
\end{eqnarray}
to take into account possible non-Regge behavior for $\hat{t}\sim\hat{s}/2$, i.e. for
small rapidity separation $\Delta Y$ between the final pions. In this paper we do not use such a ``mixed'' method, and
consider these two cases (``bare'' or ``reg\-gei\-zed'' virtual pion propagator) separately, especially to see the relevance
of the Regge approach in the kinematical region $\hat{t}\sim\hat{s}/2\sim 1$~GeV$^2$. The Regge model 
really does not work in this area or it needs to be modified (as was done, for example, in Refs.~\cite{CEDPw4}-\cite{CEDPw6}, 
with empirical formulas or additional assumptions), and 
also taking into account, for example, possible significant contribution of the background integral 
in the Sommerfeld--Watson transform (see~\cite{CollinsReggeModel}) or 
taking Legendre polynomials of order $\alpha_{\pi}(\hat{t})$ instead of 
the classical Regge term (see~(\ref{eq:PionPropagatorR})) in the ``reggeized'' virtual pion 
propagator. All of this should be verified in 
future research.
  
\begin{figure}[hbt!]
\begin{center}
\includegraphics[width=0.45\textwidth]{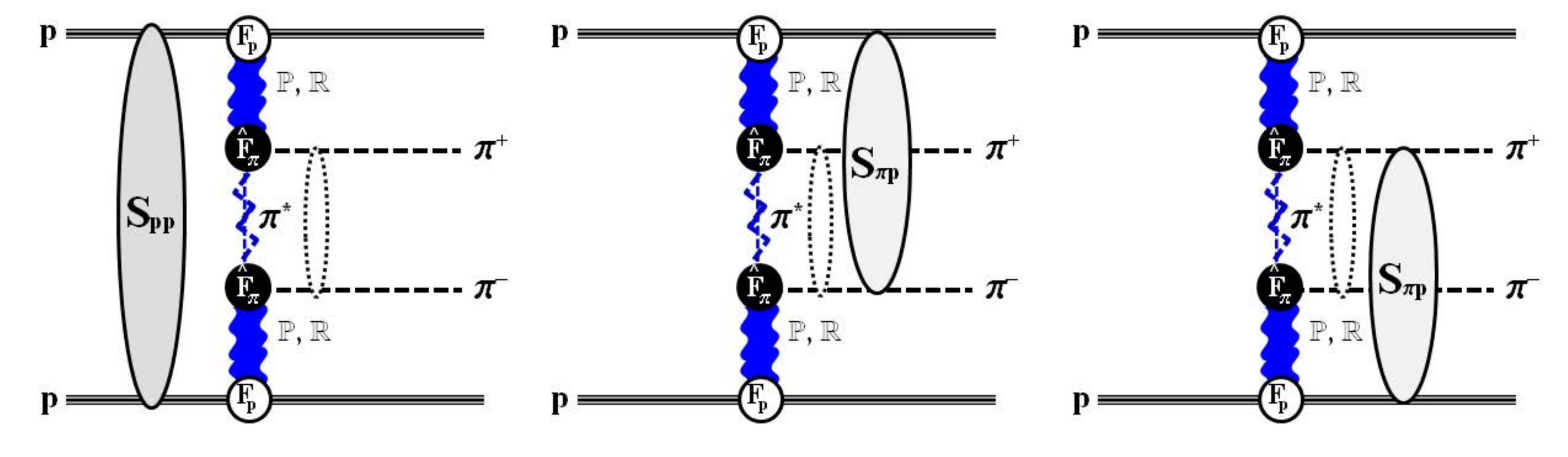}
\caption{\label{fig2:LIB} Amplitude of the process of double pion LM CEDP $p+p\to p+\pi^++\pi^-+p$ in the approach~\cite{CEDPw4}-\cite{CEDPw6}. The central part of the diagram is the Born amplitude (with Pomeron and two reggeons). A mixture of reggeized and bare off-shell pion propagators is shown as a dashed zigzag line plus dashed straight line. Proton--proton rescattering is depicted as the  $S_{pp}$-blob, and pion--proton rescattering corrections are also shown as $S_{\pi p}$-blobs. The off-shell pion form factor is presented as a black circle.}
\end{center}
\end{figure}

In Refs.~\cite{CEDPw4}-\cite{CEDPw6} the authors do not introduce
``enhanced'' corrections, but they take into account pion--proton interactions
in the final state (see Fig.~\ref{fig2:LIB}). There still is an issue
in this approach, since they mix a partially Regge approach with continuous complex spin
and a model with fixed ``Po\-me\-ron spin'' (exactly vector or tensor Pomerons). It may be convenient for
calculations and gives results close to reality, but we have no clear physical 
explanation. On the one hand we have collision of two particles with fixed 
spin (1 or 2), and on the other hand we use the Regge
expression, where the spin is replaced by the complex Regge trajectory.

\begin{figure}[hbt!]
\begin{center}
\includegraphics[width=0.49\textwidth]{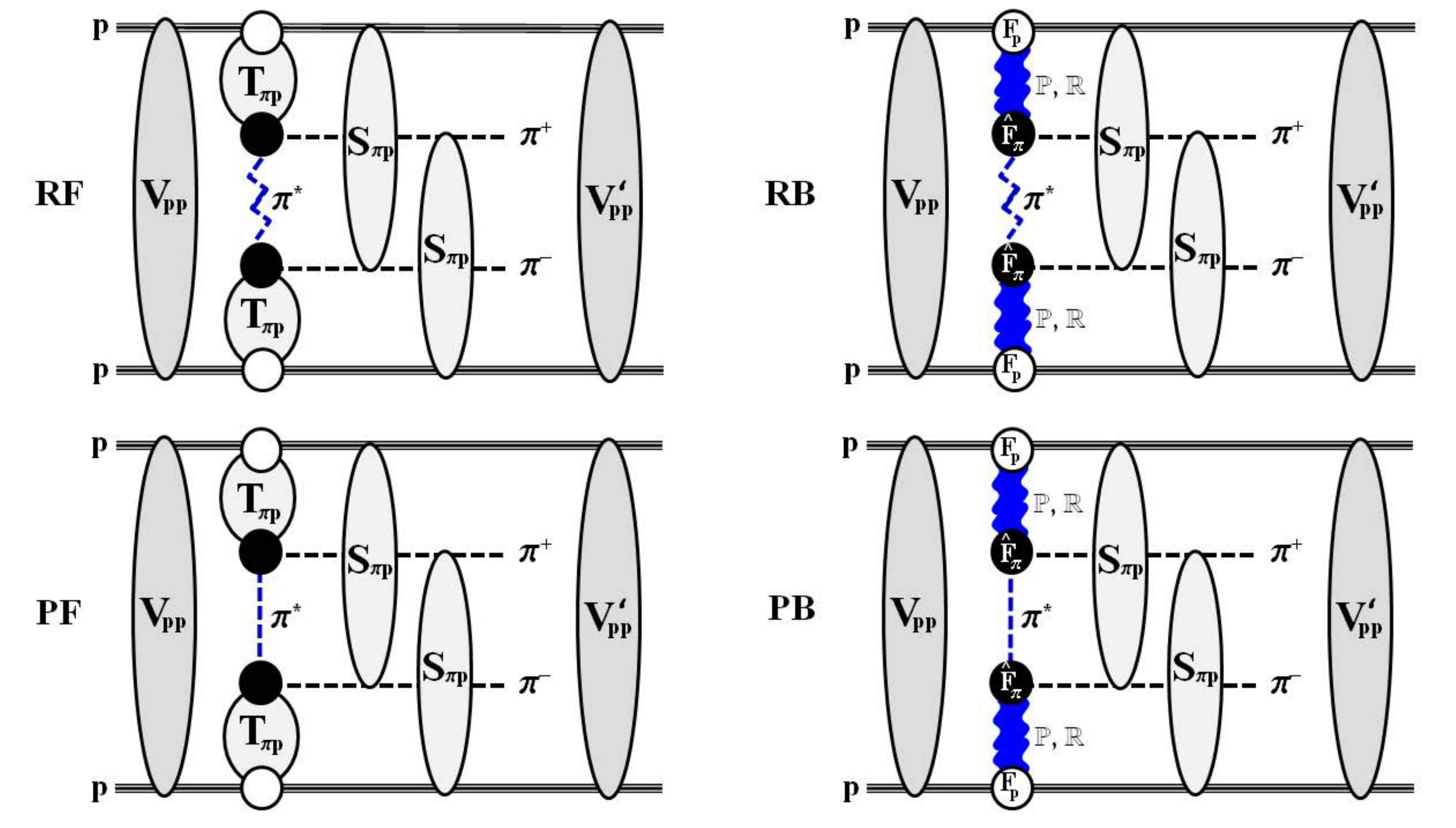}
\caption{\label{fig3:MY4CASES} Amplitude of the process of double pion LM CEDP $p+p\to p+\pi^++\pi^-+p$ in the Regge-eikonal approach for different cases. PB: central part of the diagram is the Born amplitude (with Pomeron and two reggeons) plus an off-shell bare pion propagator depicted as a dashed straight line. RB: the same as PB, but with the reggeized off-shell pion propagator depicted as a dahshed zigzag line. PF: central part of the diagram contains full eikonalized
pion--proton amplitudes plus off-shell bare pion propagator depicted as a dashed straight line. RF: the same as before but with reggeized off-shell pion propagator depicted as a dashed zigzag line. Proton--proton rescatterings in the initial and final states are depicted as $V_{pp}$- and $V'_{pp}$-blobs, respectively, and pion--proton rescattering corrections are also shown as $S_{\pi p}$-blobs. The off-shell pion form factor is presented as a black circle.}
\end{center}
\end{figure}

In this article we consider several cases, depicted
in Fig.~\ref{fig3:MY4CASES}, and we show how they describe
the data from the ISR~\cite{ISRdata1},\cite{ISRdata2},
STAR~\cite{STARdata1},\cite{STARdata2}, CDF~\cite{CDFdata1},\cite{CDFdata2},
CMS~\cite{CMSdata1},\cite{CMSdata2} collaborations.

In the first part of the present work we introduce the framework
for calculations of double pion LM CEDP (kinematics, amplitudes, differential
cross-sections) in the Regge-eikonal approach.

In the second part we analyze the experimental data on the process
at different energies, find the best approach and make some predictions for
LHC experiments.

In the final part we discuss the possibilities to extract
Pomeron--Pomeron cross-sections from the data and ana\-lyze
the present situation. Also we show some nuances
of the calculations, which we should take into account (elastic amplitudes for virtual 
particles, off-shell pion form factor, pion--pion elastic
amplitude at low energies, and nonlinearity of the pion trajectory).
  
\section{General framework for calculations of LM CEDP}

LM CEDP is the first exclusive two to four process which is driven by the Pomeron--Pomeron fusion subprocess. That is why it serves as a basic background for LM CEDP of resonances like $f_0(980)$, $f_2(1270)$. At the moment, for low central pion--pion masses (less than $\sim 3$~GeV), it is a huge problem to use a  perturbative approach, that is why we apply the Regge-eikonal method for all the calculations. For proton--proton and proton--pion elastic amplitudes we use the model of~\cite{godizovpp}, \cite{godizovpip}, which describes all the available experimental data on elastic scattering.

\subsection{Components of the framework}

LM CEDP process can be calculated in the following scheme (see Fig.~\ref{fig3:MY4CASES}):
\begin{enumerate}

\item
We calculate the primary amplitude of the process, which is depicted as the central part of diagrams in Fig.~\ref{fig3:MY4CASES}. Here we
consider four cases to show that only one of them gives the best description of the data on this process. The case PB 
represents the Born term (the letter P means 
that we use the usual ``bare'' virtual pion propagator $1/(\hat{t}-m_{\pi}^2)$, and B 
means the Born term~(\ref{eq:godizovpip1}) in each shoulder 
or pion--proton elastic subamplitudes of the central primary amplitude). The expressions 
of pion--proton elastic subamplitudes can be found in Appendix~B.

The case RB is similar to the previous one, and the letter R (instead of P) means 
that the bare off-shell pion propagator is replaced by the reggeized one
\begin{eqnarray}
{\cal P}_{\pi}(\hat{s},\hat{t})&=&
\left(
\mathrm{ctg}\frac{\pi\alpha_{\pi}(\hat{t})}{2}-\mathrm{i}
\right)\cdot\nonumber\\
&\cdot&\frac{\pi\alpha_{\pi}^{\prime}}{2\phantom{\mbox{\large{I}}^2}\!\!\!\!\mathrm{\Gamma}(1+\alpha_{\pi}(\hat{t}))}
\left(
\frac{\hat{s}}{s_0}
\right)^{\alpha_{\pi}(\hat{t})},
\label{eq:PionPropagatorR}
\end{eqnarray}
where $\hat{s}$ is the di-pion mass squared and $\hat{t}$ is the square of the momentum transfer between a  Pomeron and a pion in the Pomeron--Pomeron fusion process (see Appendix~A for details). 

The cases PF (RF) can be obtained from PB (RB) if we replace the Born pion--proton elastic amplitudes to full eikonalized 
expressions (which is reflected in the replacement of B to F in the notation for these cases), which could be found in Appendix~B.

\item
After the calculation of the primary LM CEDP amplitude we have to take into account all possible corrections in proton--proton and proton--pion elastic channels due to the unitarization procedure (the so called ``soft survival probability'' or ``rescattering corrections''), which are depicted as $V_{pp}$, $V'_{pp}$ and $S_{\pi p}$ blobs in Fig.~\ref{fig3:MY4CASES}. For the proton--proton and proton--pion elastic amplitudes we use the model of~\cite{godizovpp}, \cite{godizovpip} (see Appendix~B). A possible final pion--pion interaction is not 
shown in Fig.~\ref{fig3:MY4CASES}, since we neglect it in the present calculations. RB and PB cases of Fig.~\ref{fig3:MY4CASES} are similar to the one in Fig.~\ref{fig2:LIB}.

\end{enumerate}

In this article we do not consider the so called ``enhanced'' corrections~\cite{CEDPw1}-\cite{CEDPw3}, since they give nonleading contributions in our model due to the smallness of the triple Pomeron vertex. Also we have no possible absorptive  corrections in the pion--pion final elastic channel, since the central mass is low, and also there is a lack of data on this process to define parameters of the model. Nevertheless we will consider these corrections in further work, as was done by some authors recently~\cite{pipiespagnol}, since they could play significant role for masses less than 1~GeV.

Exact kinematics of the two to four process is outlined in Appendix~A.

Here we use the model, presented in Appendix~B for example. One can use another one, which has been proved 
to describe well all the available data on proton--proton and proton--pion elastic processes. But it is difficult to find 
now more than a couple of models 
which have more or less predictable 
power (see~\cite{godizovmodelsfault} for a detailed discussion). That is why we use the model, which has been proved to be good 
in data fitting, especially in the kinematical region of our in\-te\-rest.

The final expression for the amplitude with proton--pro\-ton and pion--proton ``rescattering'' corrections can be written as
\begin{eqnarray}
&& M^U\left( \{ p \}\right) = 
\nonumber\\
&&  
=\int\int 
\frac{d^2\vec{q}}{(2\pi)^2}\frac{d^2\vec{q}^{\prime}}{(2\pi)^2}
\frac{d^2\vec{q}_1}{(2\pi)^2}\frac{d^2\vec{q}_2}{(2\pi)^2}
V_{pp}(s,q^2)
V_{pp}(s^{\prime},q^{\prime 2})\nonumber\\
&& \times\; \left[
S_{\pi^-p}(\tilde{s}_{14},q_1^2)
M_0\left( \{ \tilde{p}\} \right)
S_{\pi^+p}(\tilde{s}_{23},q_2^2)+
(3\leftrightarrow 4)
\right] \label{eq:MU}\\
&& M_0\left( \{ p \}\right)=\phantom{I^{2^{2^2}}}\nonumber\\
&&
=T^{el}_{\pi^+p}(s_{13},t_1)
{\cal P}_{\pi}(\hat{s},\hat{t})
\left[ 
\hat{F}_{\pi}\left( \hat{t}\right)
\right]^2
T^{el}_{\pi^-p}(s_{24},t_2),
\label{eq:M0}
\end{eqnarray}
where the functions are defined in~(\ref{eq:Vpp1})--(\ref{eq:Tpipexact}) of Appendix~B, and the sets of vectors are
\begin{eqnarray}
&&\{ p \}\equiv \{ p_a,p_b,p_1,p_2,p_3,p_4\},\label{eq:setp}\\
&&\{ \tilde{p}\}\equiv \{ p_a-q,p_b+q; p_1+q^{\prime}+q_1,\nonumber\\
&& \;\;\;\;\;\;\;\;\;\;\;\; p_2-q^{\prime}+q_2,p_3-q_2,p_4-q_1 \} \label{eq:setptild},
\end{eqnarray}
and
\begin{eqnarray}
\tilde{s}_{14}&=&\left( p_1+p_4+q^{\prime} \right)^2,\;
\tilde{s}_{23}=\left( p_2+p_3-q^{\prime} \right)^2,
\label{eq:invarstild}\\
s_{ij}&=&\left( p_i+p_j \right)^2,\;
t_{1,2}=\left( p_{a,b}-p_{1,2} \right)^2,
\label{eq:invars}\\
\hat{s}&=&\left( p_3+p_4 \right)^2,\;
\hat{t}=\left( p_a-p_1-p_3 \right)^2
\label{eq:invarsh}
\end{eqnarray}

The off-shell pion form factor is equal to unity on the mass shell, $\hat{t}=m_{\pi}^2$, and 
taken as exponential
\begin{equation}
\hat{F}_{\pi}=\mathrm{e}^{(\hat{t}-m_{\pi}^2)/\Lambda_{\pi}^2},
\label{eq:offshellFpi}
\end{equation}
where $\Lambda_{\pi}$ is taken from 
the fits to LM CEDP of two pions at low energies (see next section). In this paper we use only exponential form, but
it is possible to use other parametrizations (see~\cite{CEDPw1}-\cite{CEDPw6}). The xponential form shows more appropriate
results in the data fitting.

Other functions are defined in Appendix~B. Then we can use Eq.~(\ref{eq5:dcsdall}) to calculate
the differential cross-section of the process.

\subsection{Nuances of calculations}

In the next section one can see that there are some difficulties in the data fitting, which 
have also been presented in other work~\cite{CEDPw4}-\cite{CEDPw6}. In this 
section let us discuss some nuances of calculations, which could change the situation. 

We have to pay special attention to the amplitudes, where one or more external particles
are off their mass shell. The example of such an amplitude is the
pion--proton one $T_{\pi^+ p}$ ($T_{\pi^- p}$), which is the part of the CEDP 
amplitude (see~(\ref{eq:MU})). For this amplitude in the present paper we use the Regge-eikonal model with the
eikonal function in the classical Regge form. The ``off-shell'' condition for one of the pions
is taken into account by an additional phenomenological form factor 
$\hat{F}_{\pi}(\hat{t})$. But there are at least two other possibilities.

The first one was considered in~\cite{PetrovOffshell}. For the amplitude 
with one particle off-shell the formula
\begin{equation}
T^*(s,b)=\frac{\delta^*(s,b)}{\delta(s,b)}T(s,b)=\frac{\delta^*(s,b)}{\delta(s,b)}\frac{\mathrm{e}^{2\mathrm{i}\delta(s,b)}-1}{2\mathrm{i}}
\label{eq:TstarVAP}
\end{equation}
was used. In our case 
\begin{eqnarray}
&& \delta(s,b) = \delta_{\pi p}(s,b; m_{\pi}^2,m_{\pi}^2,m_p^2,m_p^2),\nonumber\\ 
&& \delta^*(s,b) = \delta_{\pi p}^*(s,b; \hat{t},m_{\pi}^2,m_p^2,m_p^2)\nonumber\\
&& \left.\delta_{\pi p}=\delta^*_{\pi p}\right|_{\hat{t}\to m_{\pi}^2}.
\label{eq:deltasVAP}
\end{eqnarray}
$\delta_{\pi p}$ is the eikonal function (see~(\ref{eq:elamplitudes})). This is similar to the introduction of the additional form factor, but in a more consistent way, which takes into account the 
unitarity condition.

\begin{figure}[hbt!]
\begin{center}
\includegraphics[width=0.45\textwidth]{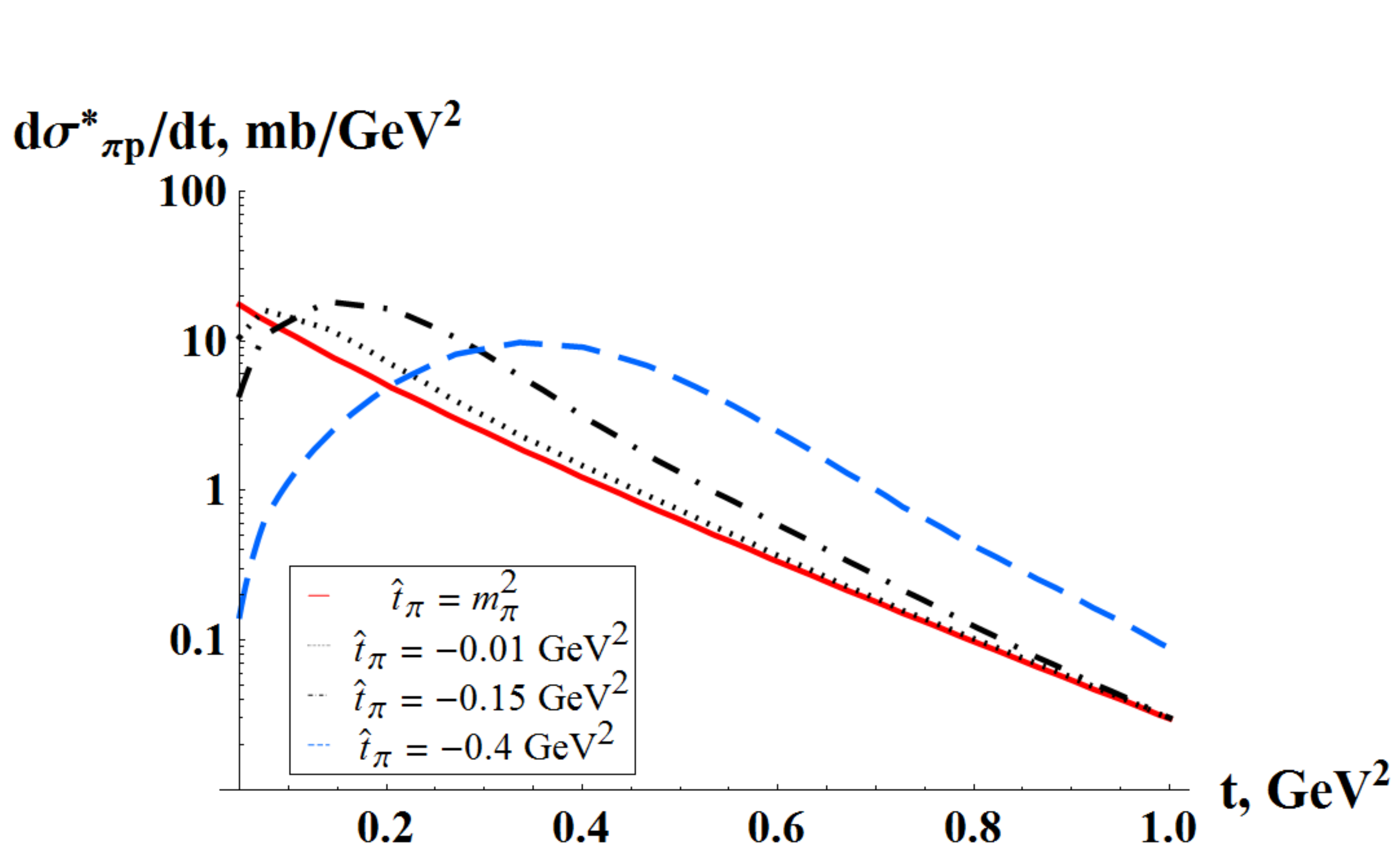}
\caption{\label{fig:offshellpip} Pion--proton on-shell and off-shell elastic differential cross-section (in the
	model of conserved meson currents presented in Appendix~C) for different pion virtualities $\hat{t}_{\pi}$: $m_{\pi}^2$(on-shell),$-0.01$~$\mathrm{GeV}^2$,$-0.15$~$\mathrm{GeV}^2$,$-0.4$~$\mathrm{GeV}^2$ in the covariant approach with conserved currents~(\ref{UneqMass}).}
\end{center}
\end{figure}   

The second one arises from the covariant reggeization
method, which is considered in Appendix~C. For the case of conserved hadronic currents
we have the definite structure in the Legendre function~(\ref{UneqMass}), which is transformed in a natural way to the case of the off-shell
amplitude. But in this case the off-shell amplitude shows a specific behavior at low t values (see Fig.~\ref{fig:offshellpip} and~\cite{myCEDP2} for details). As 
was shown in~\cite{myCEDP2}, unitarity corrections can mask this behavior. To check this we need to make all the calculations and fitting of the data for
the process $p+p\to p+\pi+\pi+p$, but with an amplitude like~(\ref{UneqMass}) instead of~(\ref{eq:Vpip2}). This will be done in further work on 
the subject.

In the present calculations we use the linear pion trajectory $0.7(\hat{t}-m_{\pi}^2)$. The nonlinear case was also
verified, and the difference in the final result is not significant.

\section{Data from hadron colliders versus results of calculations}
\label{sec:data}

Our basic task is to extract the fundamental information on the interaction of hadrons from
different cross-sections (``diffractive patterns''):
\begin{itemize}
\item from t-distributions we can obtain the size and shape of the interaction region;
\item the distribution on the azimuthal angle between the final protons gives the  quantum numbers of the
produced system (see~\cite{myCEDP2},\cite{myWA102} and the references therein);
\item from $M_c$ (here $M_c=M_{\pi\pi}$) dependence and its influence on the t-dependence we can draw some conclusions about
the interaction at different space-time scales and the interrelation between them.
\end{itemize}

The process $p+p\to p+\pi+\pi+p$  is the first ``standard candle'', which we can use to estimate other LM CEDP processes, like
resonance production~\cite{myWA102},\cite{GodizovResonances}. In this section we consider
the experimental data on the process and its description for different model cases.

\subsection{STAR collaboration data versus model cases}

\begin{figure}[h!]
\begin{center}
a)\includegraphics[width=0.35\textwidth]{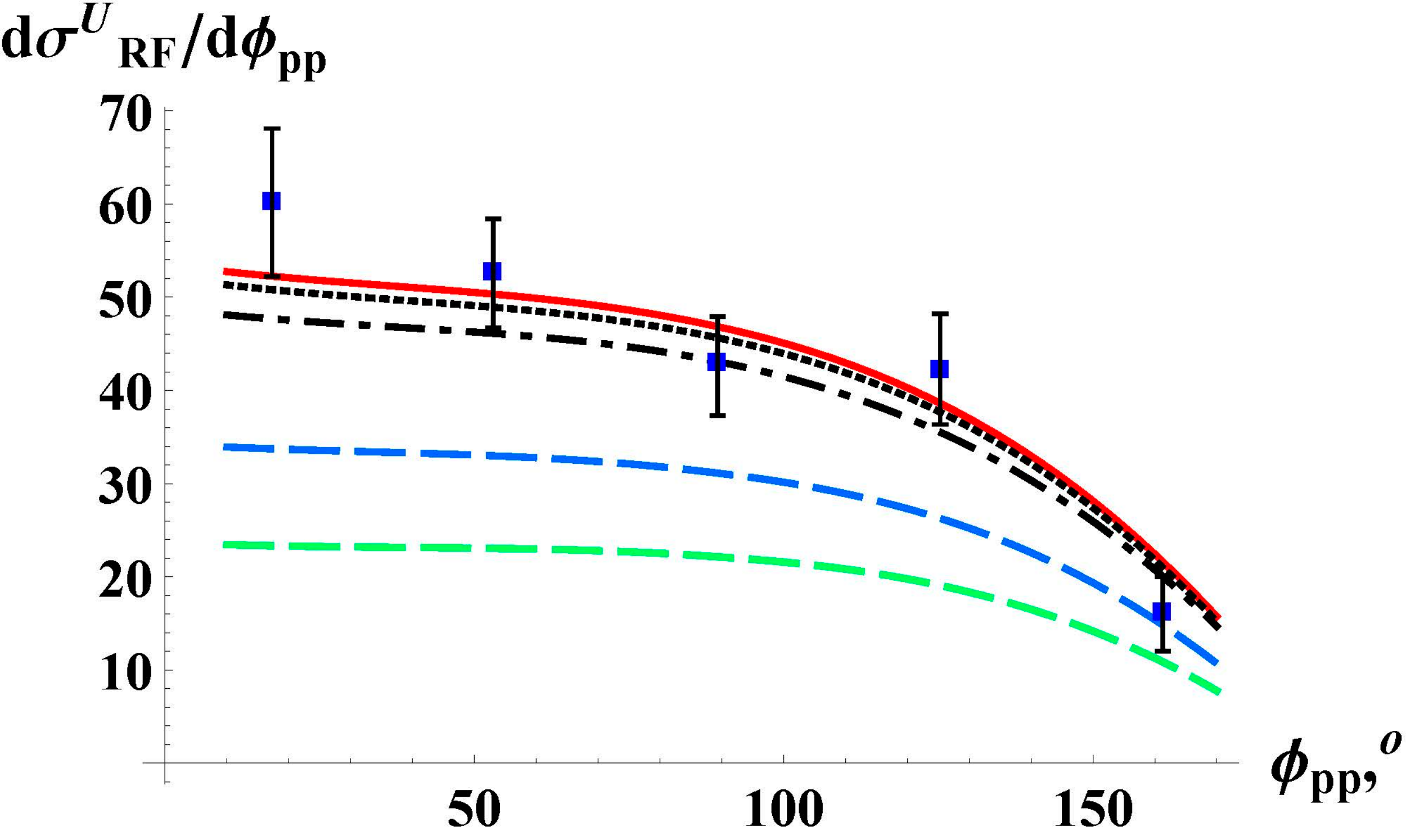}\\
b)\includegraphics[width=0.35\textwidth]{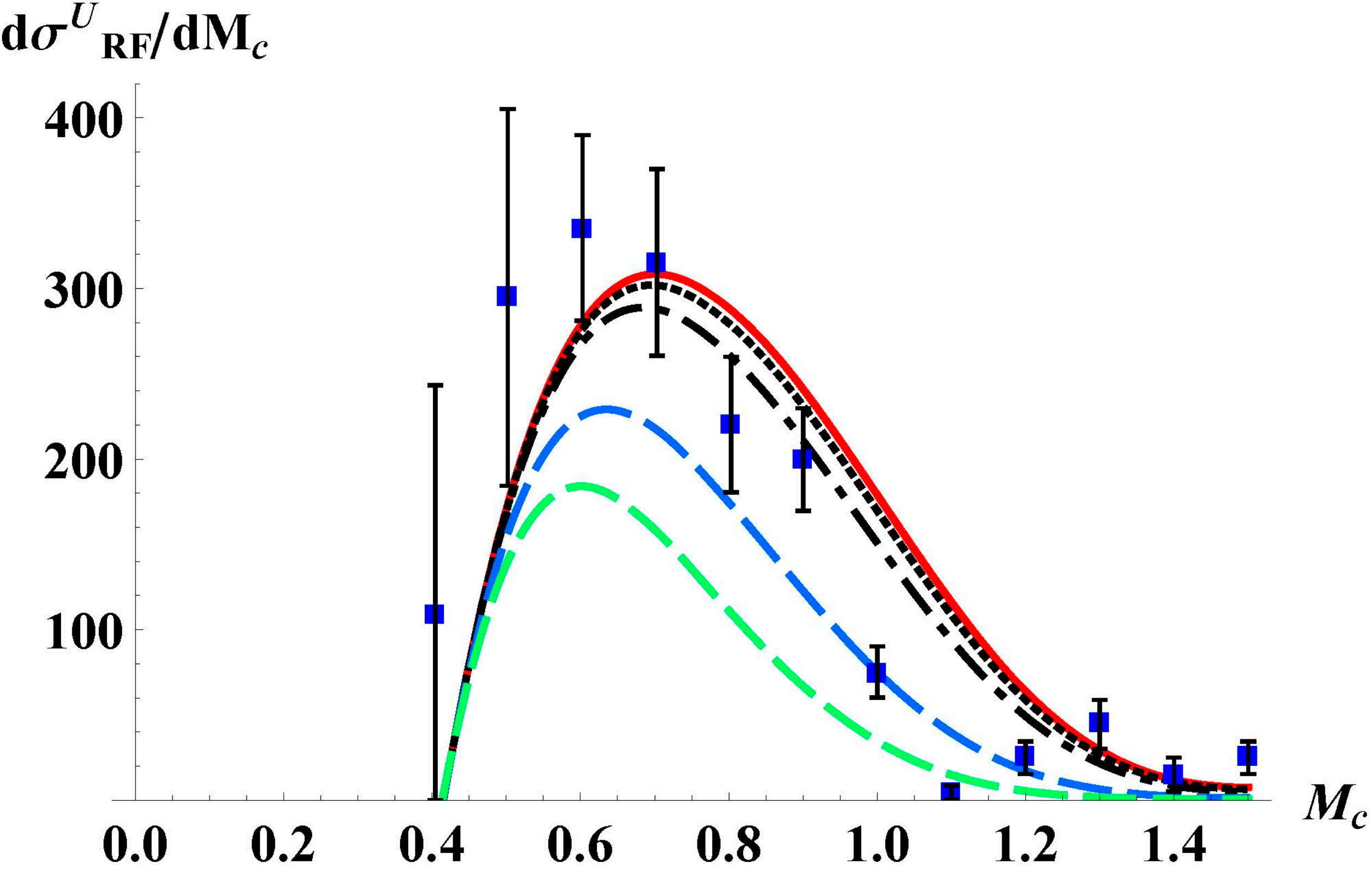}\\
c)\includegraphics[width=0.35\textwidth]{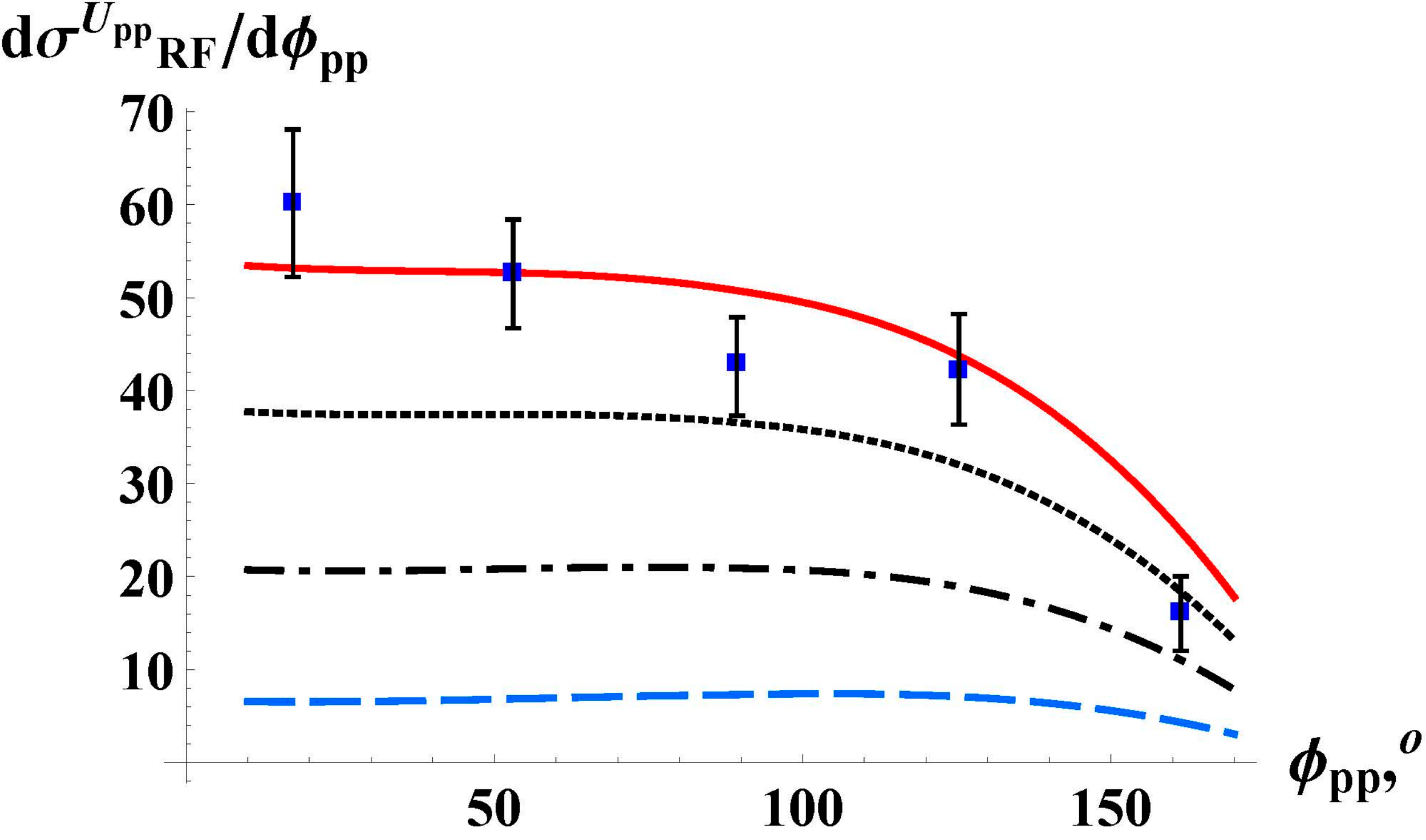}\\
d)\includegraphics[width=0.35\textwidth]{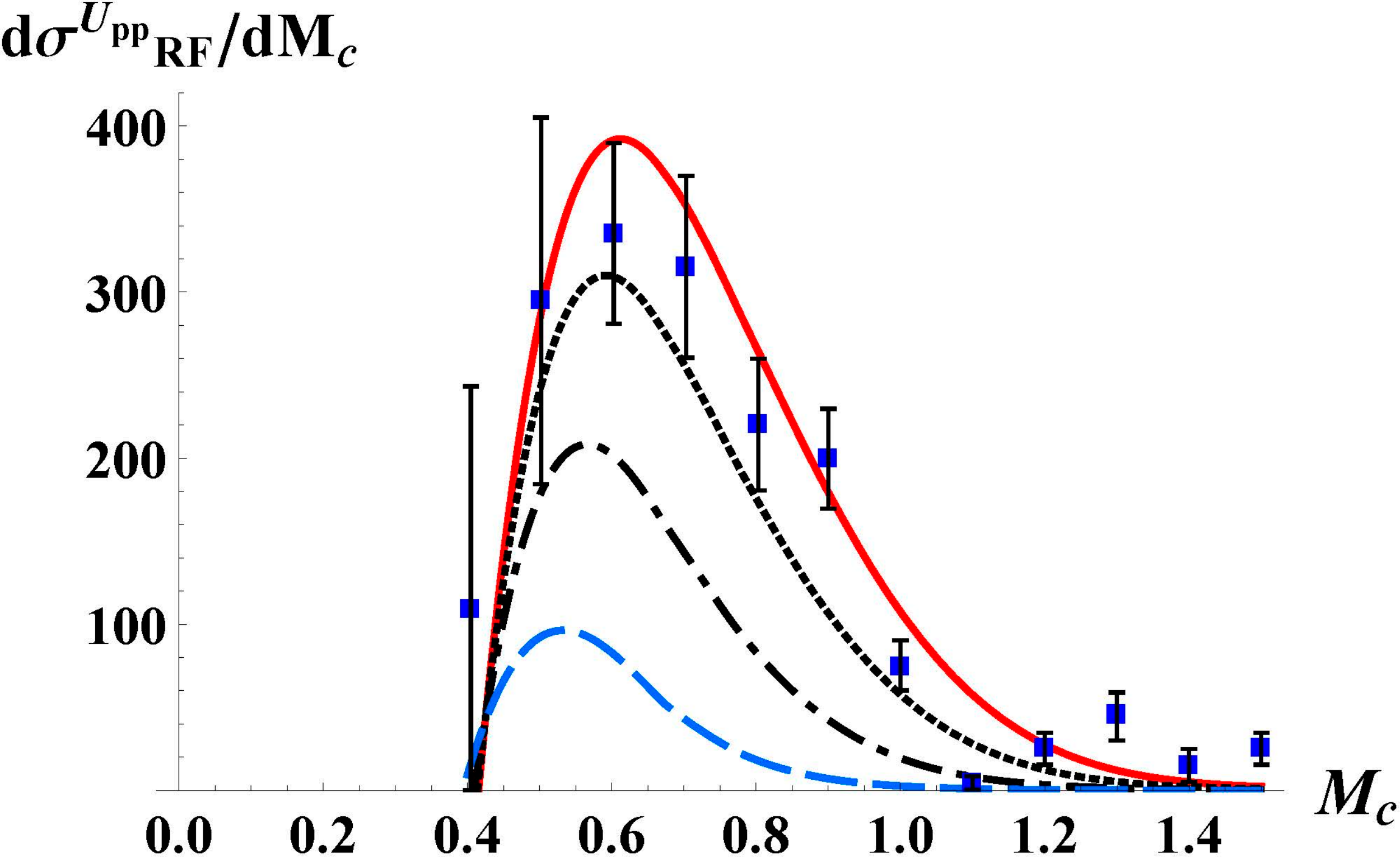}
\caption{\label{fig:starRF} RF case of the model (see Fig.~\ref{fig3:MY4CASES}). The data on the process $p+p\to p+\pi^++\pi^-+p$ 
at $\sqrt{s}=200$~GeV, $|\eta_{\pi}|<1$, $|\eta_{\pi\pi}|<2$, $p_{T\pi}>0.15$~GeV, $0.005<-t_{1,2}<0.03$~GeV$^2$, (STAR collaboration~\cite{STARdata1},\cite{STARdata2}).  {\bf a, b} The result 
with all p p and $\pi$ p rescattering corrections,  {\bf c, d} show the result, when we try to fit the data by formulas without $\pi$ p rescattering terms. Curves from 
up to down correspond to different values of the parameter $\Lambda_{\pi}$ in the off-shell 
pion form factor~(\ref{eq:offshellFpi}): {\bf a, b} $\Lambda_{\pi}=5,4,3,1.6,1.2$~GeV, (c),(d) $\Lambda_{\pi}=1.2,1,0.8,0.6$~GeV.
}
\end{center}
\end{figure}

\begin{figure}[h!]
\begin{center}
a)\includegraphics[width=0.35\textwidth]{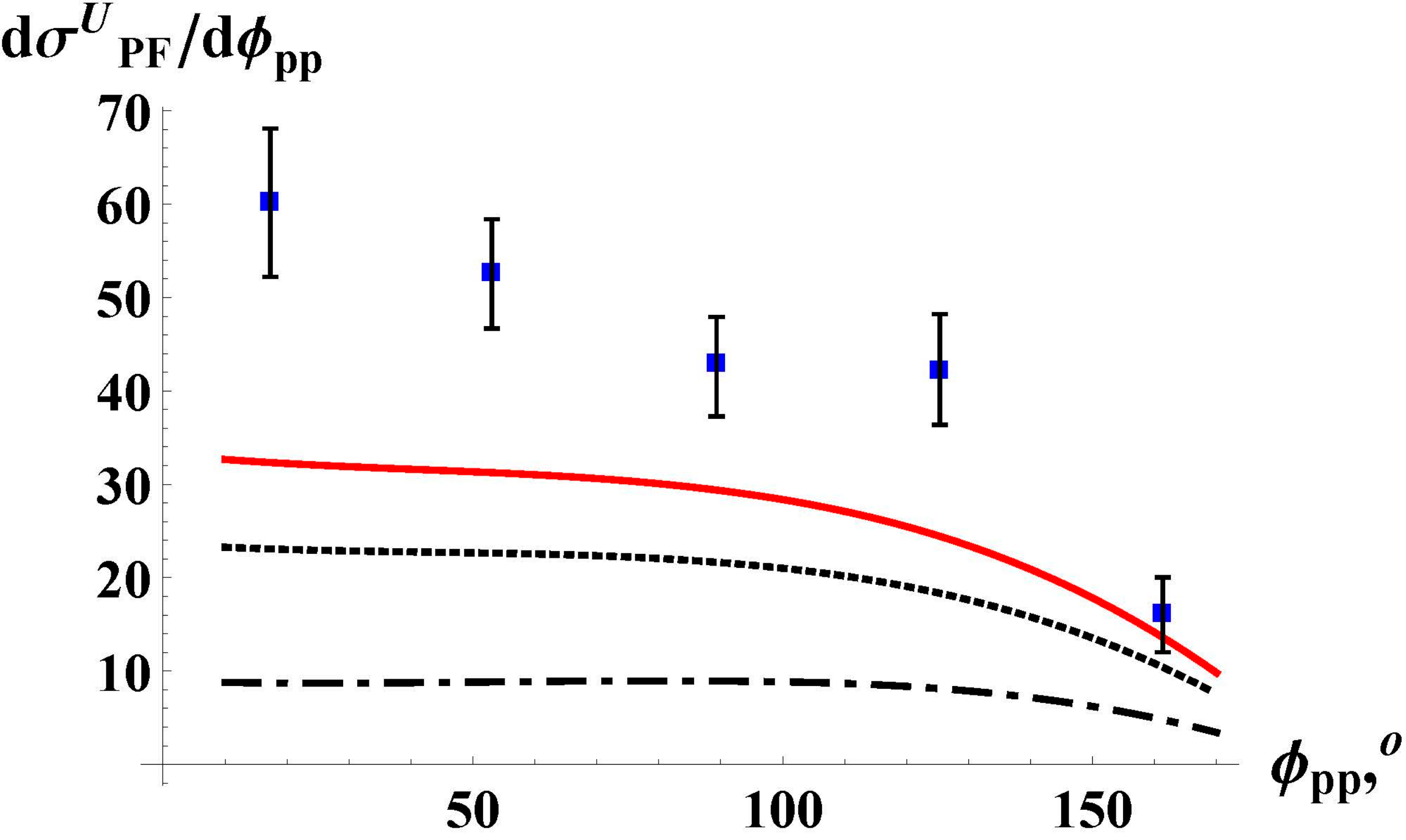}\\
b)\includegraphics[width=0.35\textwidth]{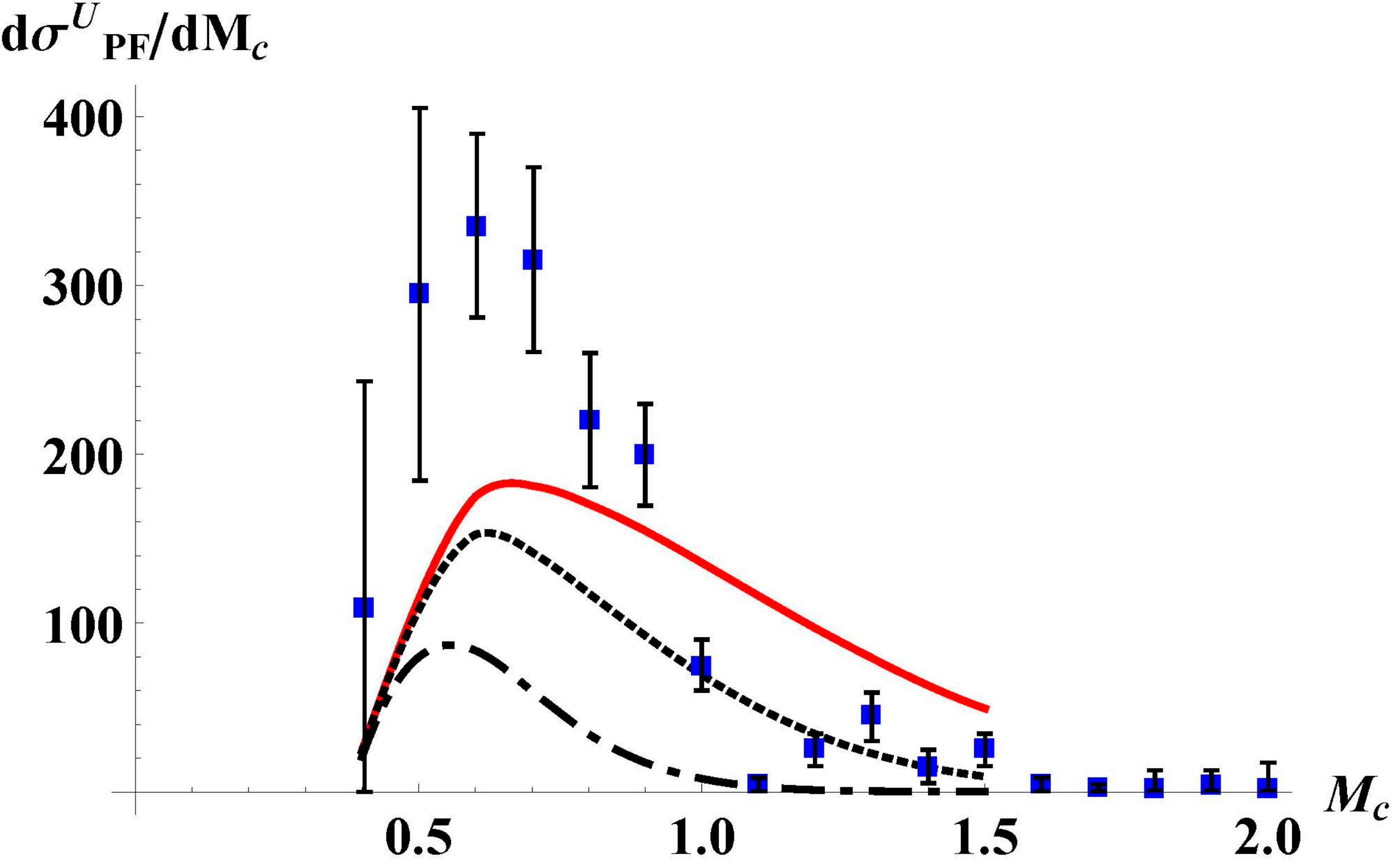}\\
c)\includegraphics[width=0.35\textwidth]{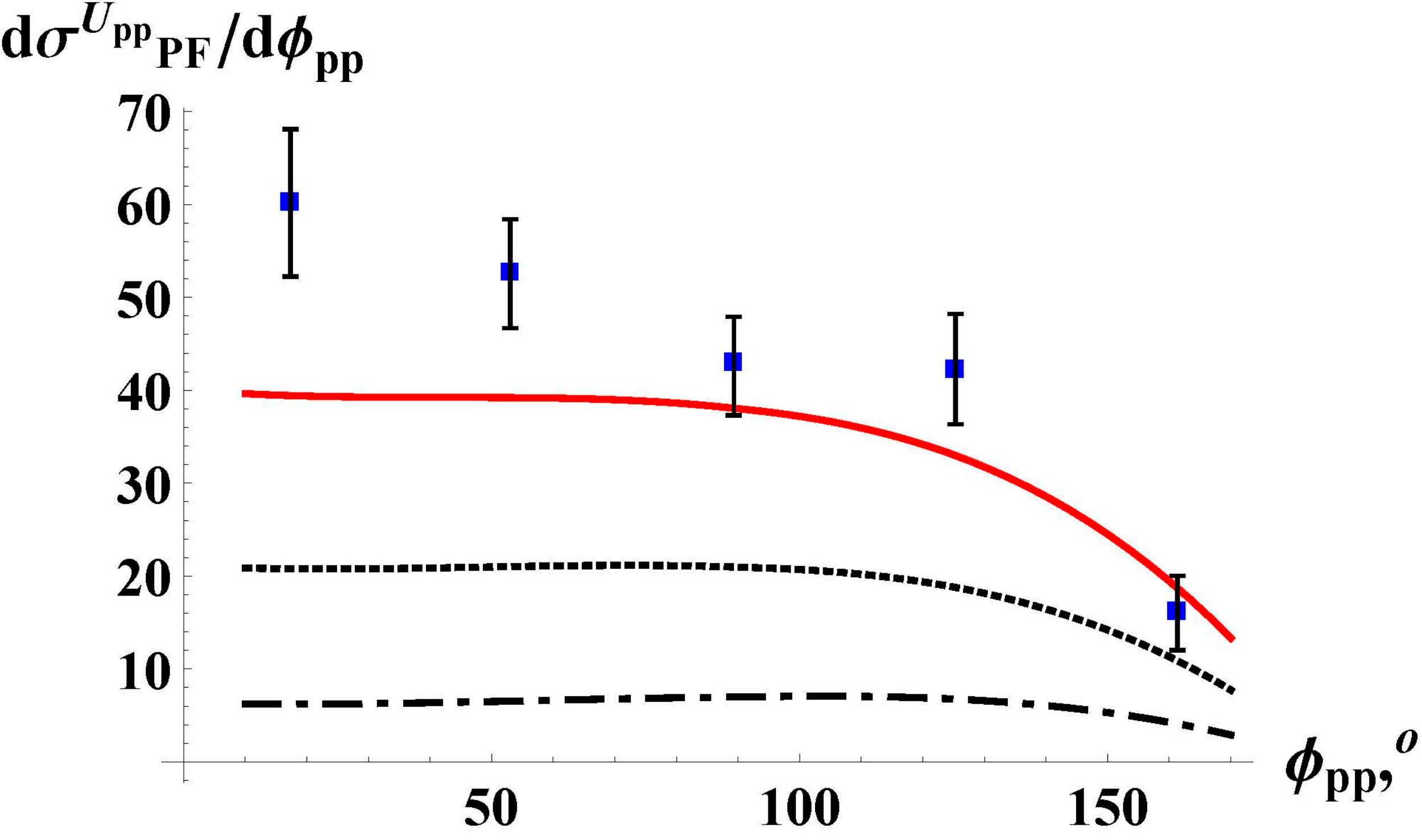}\\
d)\includegraphics[width=0.35\textwidth]{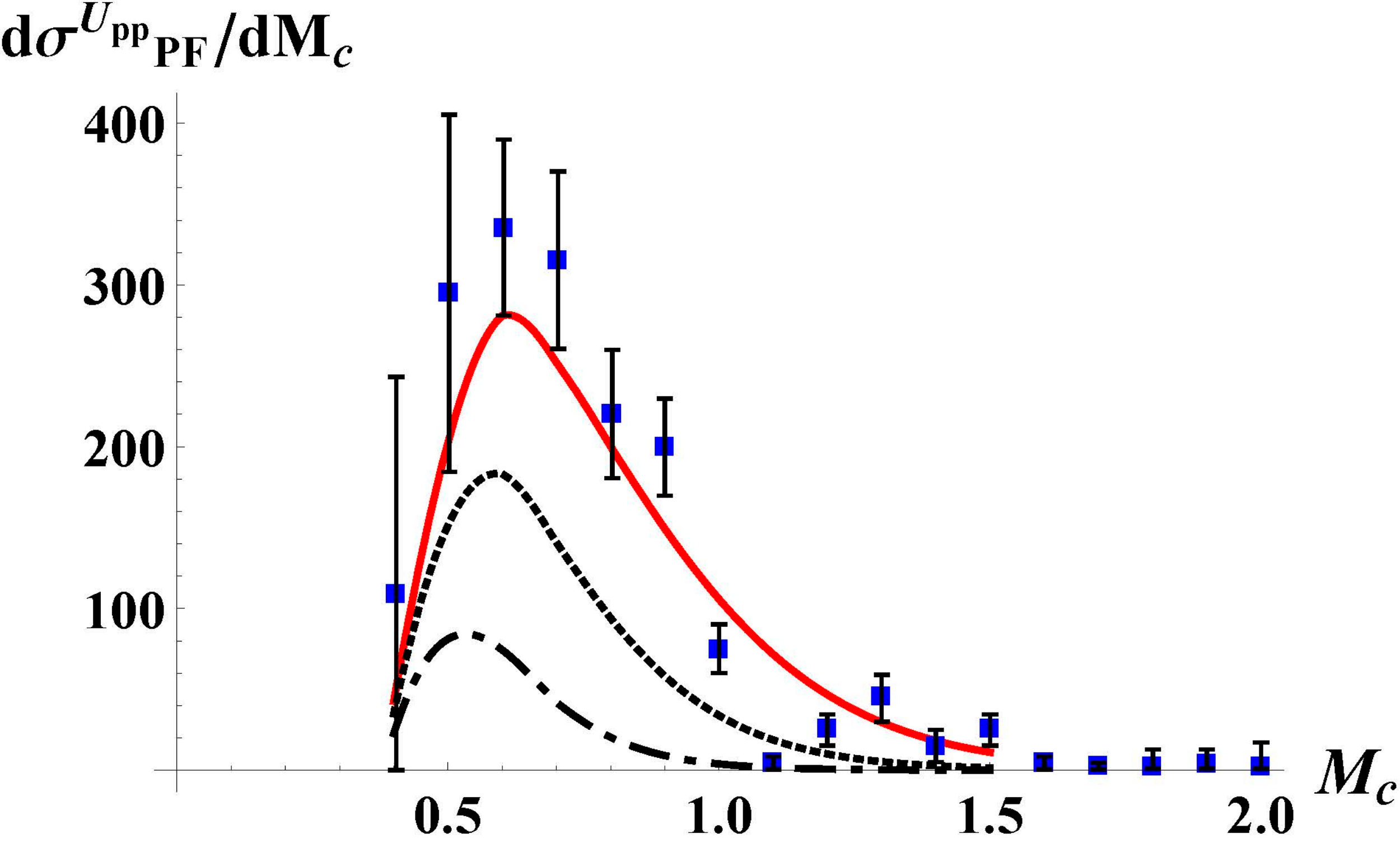}
\caption{\label{fig:starPF} PF case of the model (see Fig.~\ref{fig3:MY4CASES}). The data on the process $p+p\to p+\pi^++\pi^-+p$ 
at $\sqrt{s}=200$~GeV, $|\eta_{\pi}|<1$, $|\eta_{\pi\pi}|<2$, $p_{T\pi}>0.15$~GeV, $0.005<-t_{1,2}<0.03$~GeV$^2$, (STAR collaboration~\cite{STARdata1},\cite{STARdata2}). 
{\bf a, b} The result 
with all p p and $\pi$ p rescattering corrections,  
{\bf c, d} show the result when we try to fit the data by formulas without $\pi$ p rescattering terms. Curves from 
up to down correspond to different values of the parameter $\Lambda_{\pi}$ in the off-shell 
pion form factor~(\ref{eq:offshellFpi}): {\bf a, b} $\Lambda_{\pi}=1.6,1.2,0.8$~GeV, {\bf c, d} $\Lambda_{\pi}=1,0.8,0.6$~GeV.
}
\end{center}
\end{figure}

\begin{figure}[h!]
\begin{center}
a)\includegraphics[width=0.35\textwidth]{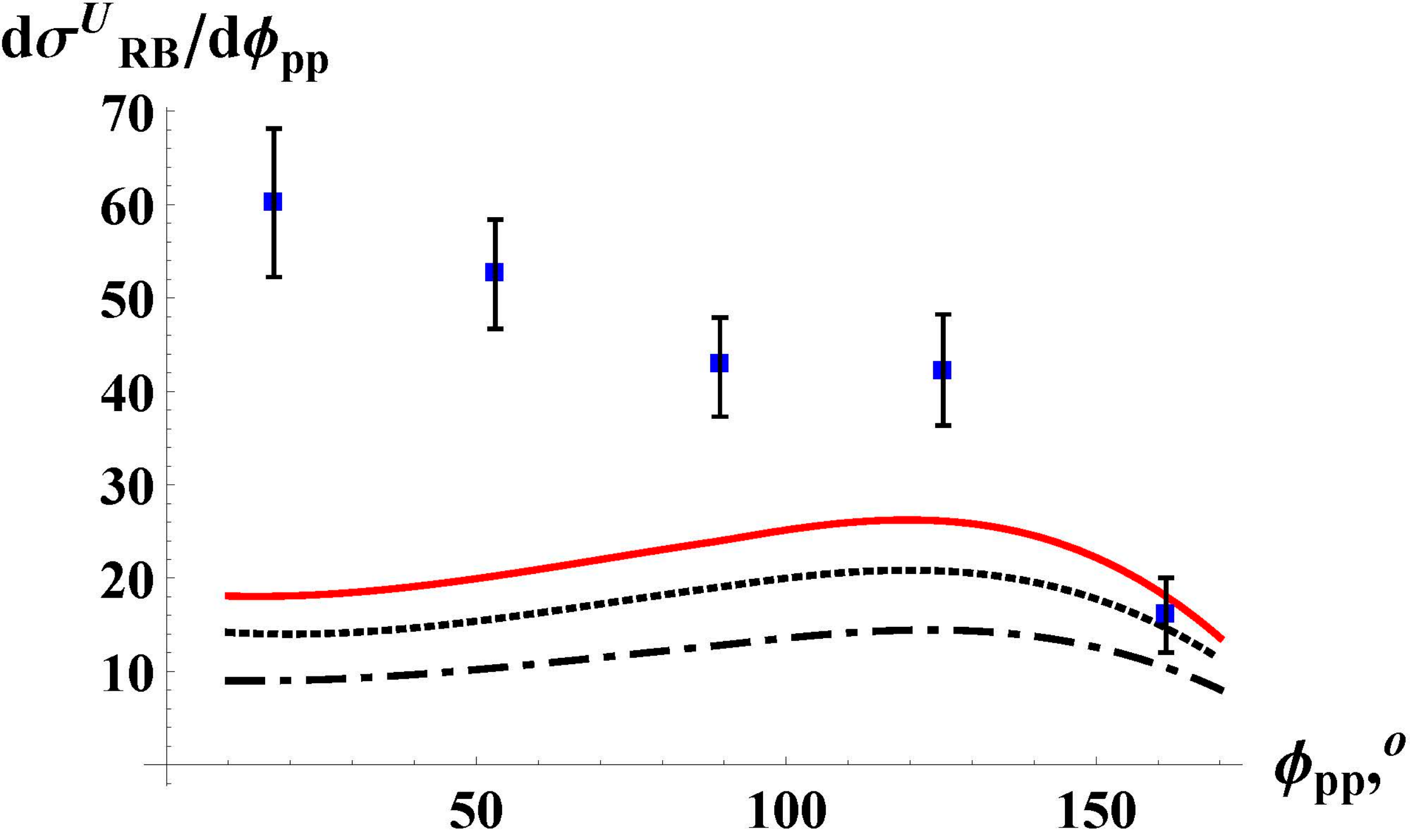}\\
b)\includegraphics[width=0.35\textwidth]{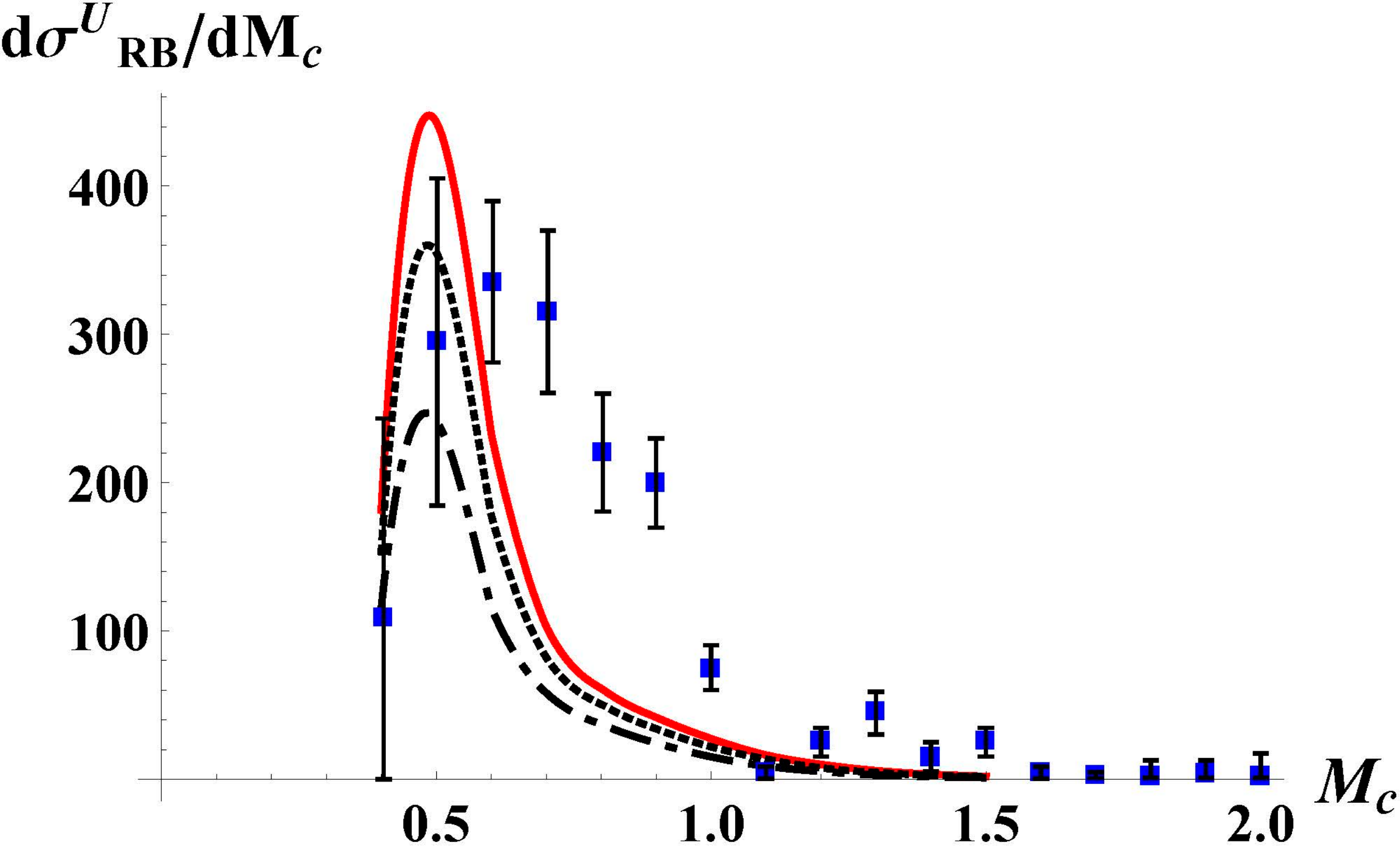}\\
c)\includegraphics[width=0.35\textwidth]{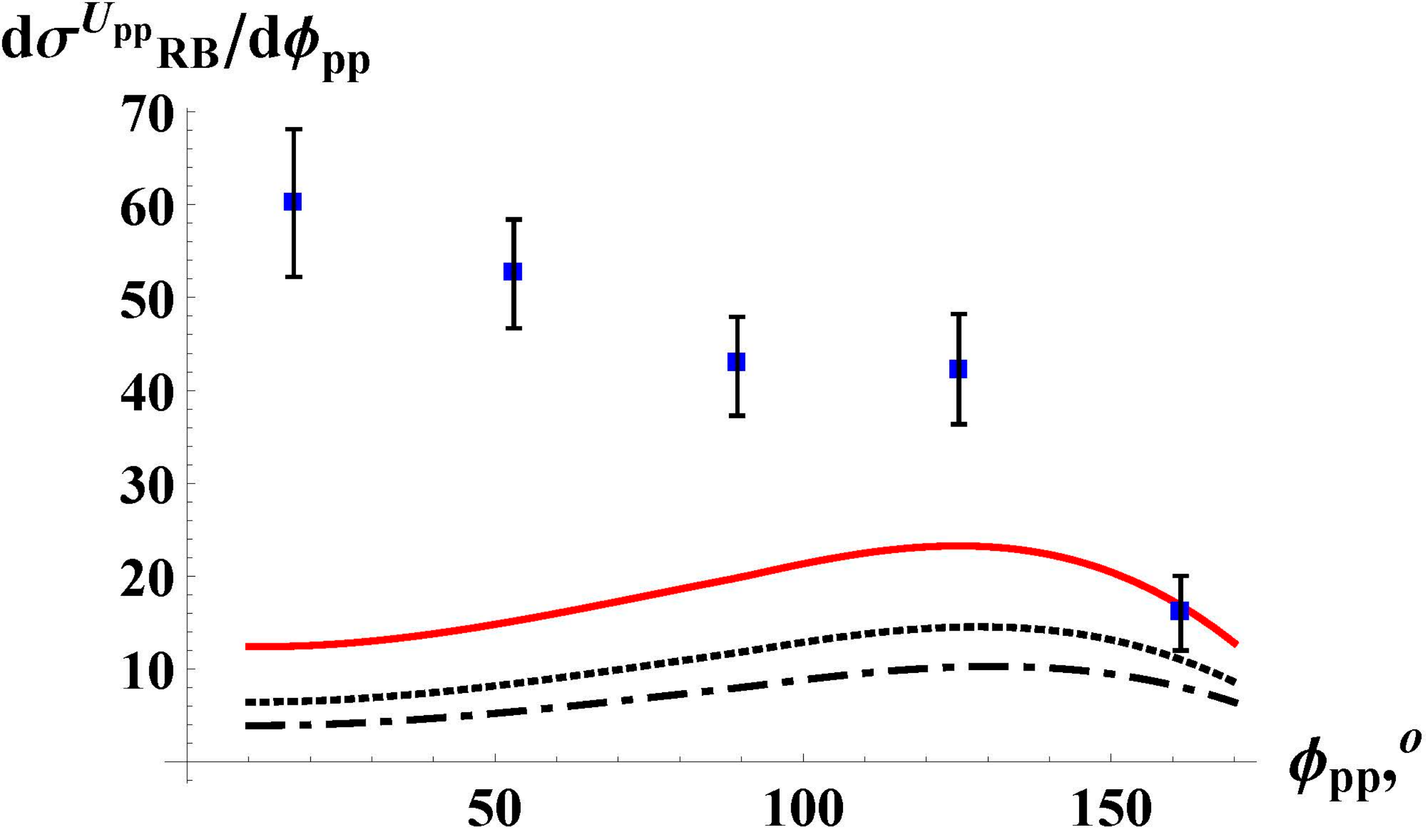}\\
d)\includegraphics[width=0.35\textwidth]{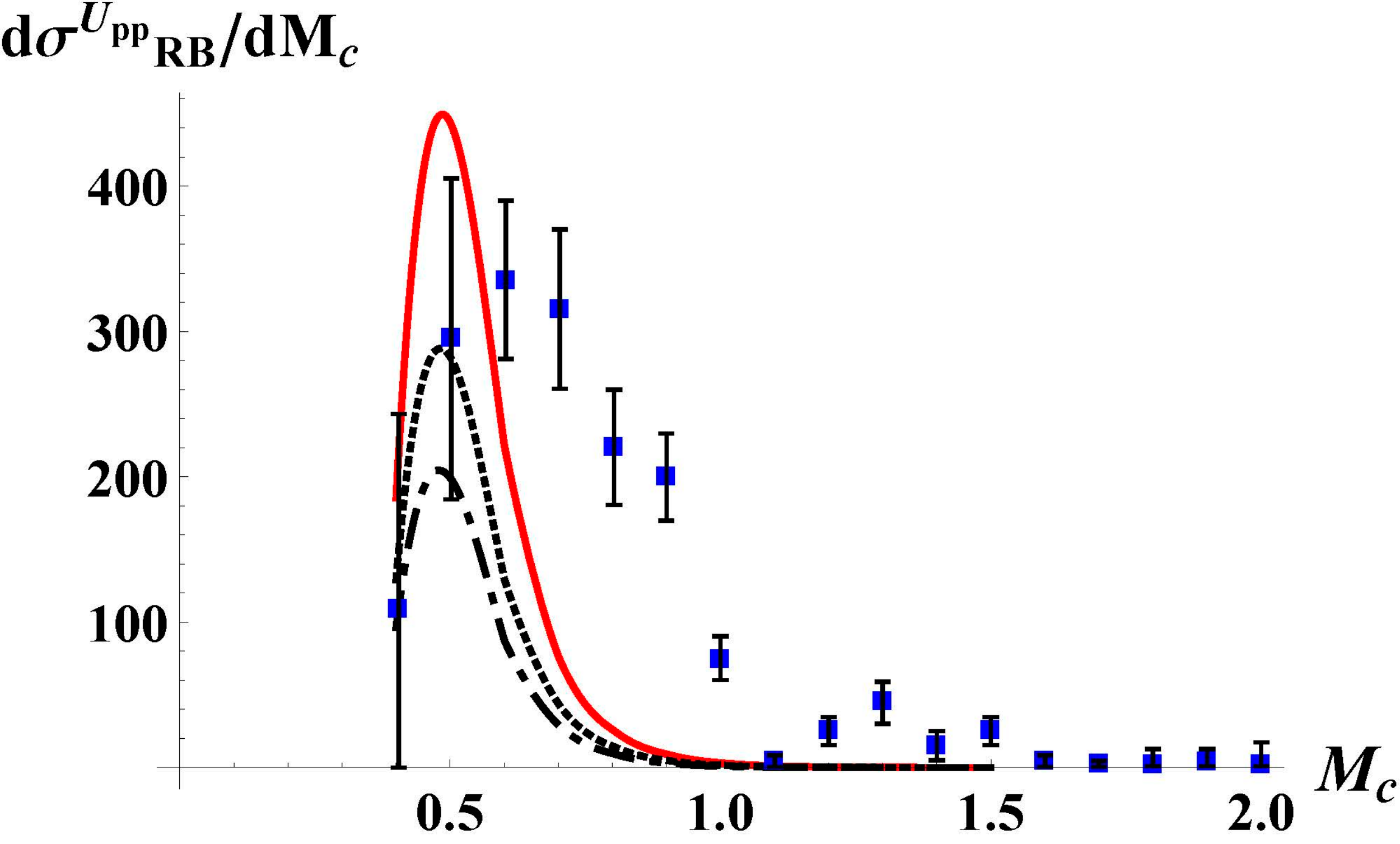}
\caption{\label{fig:starRB} RB case of the model (see Fig.~\ref{fig3:MY4CASES}). The data on the process $p+p\to p+\pi^++\pi^-+p$ 
at $\sqrt{s}=200$~GeV, $|\eta_{\pi}|<1$, $|\eta_{\pi\pi}|<2$, $p_{T\pi}>0.15$~GeV, $0.005<-t_{1,2}<0.03$~GeV$^2$, (STAR collaboration~\cite{STARdata1},\cite{STARdata2}).  {\bf a, b} The result 
with all p p and $\pi$ p rescattering corrections, 
{\bf c, d} show the result when we try to fit the data by formulas without $\pi$ p rescattering terms. Curves from 
up to down correspond to different values of the parameter $\Lambda_{\pi}$ in the off-shell 
pion form factor~(\ref{eq:offshellFpi}): {\bf a, b} $\Lambda_{\pi}=0.45,0.43,0.4$~GeV, {\bf c, d} $\Lambda_{\pi}=0.4,0.37,0.35$~GeV.
}
\end{center}
\end{figure}

\begin{figure}[h!]
\begin{center}
a)\includegraphics[width=0.35\textwidth]{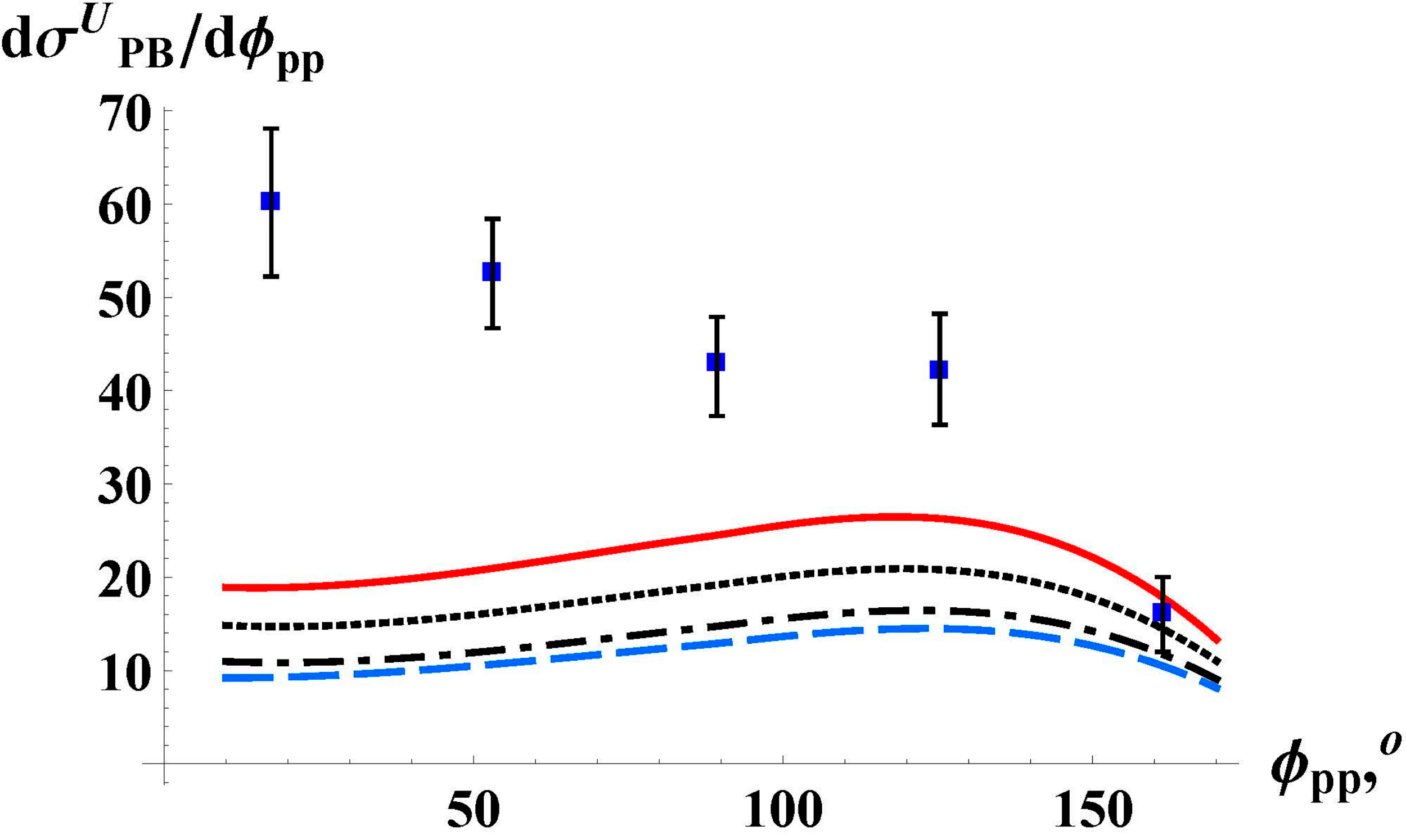}\\
b)\includegraphics[width=0.35\textwidth]{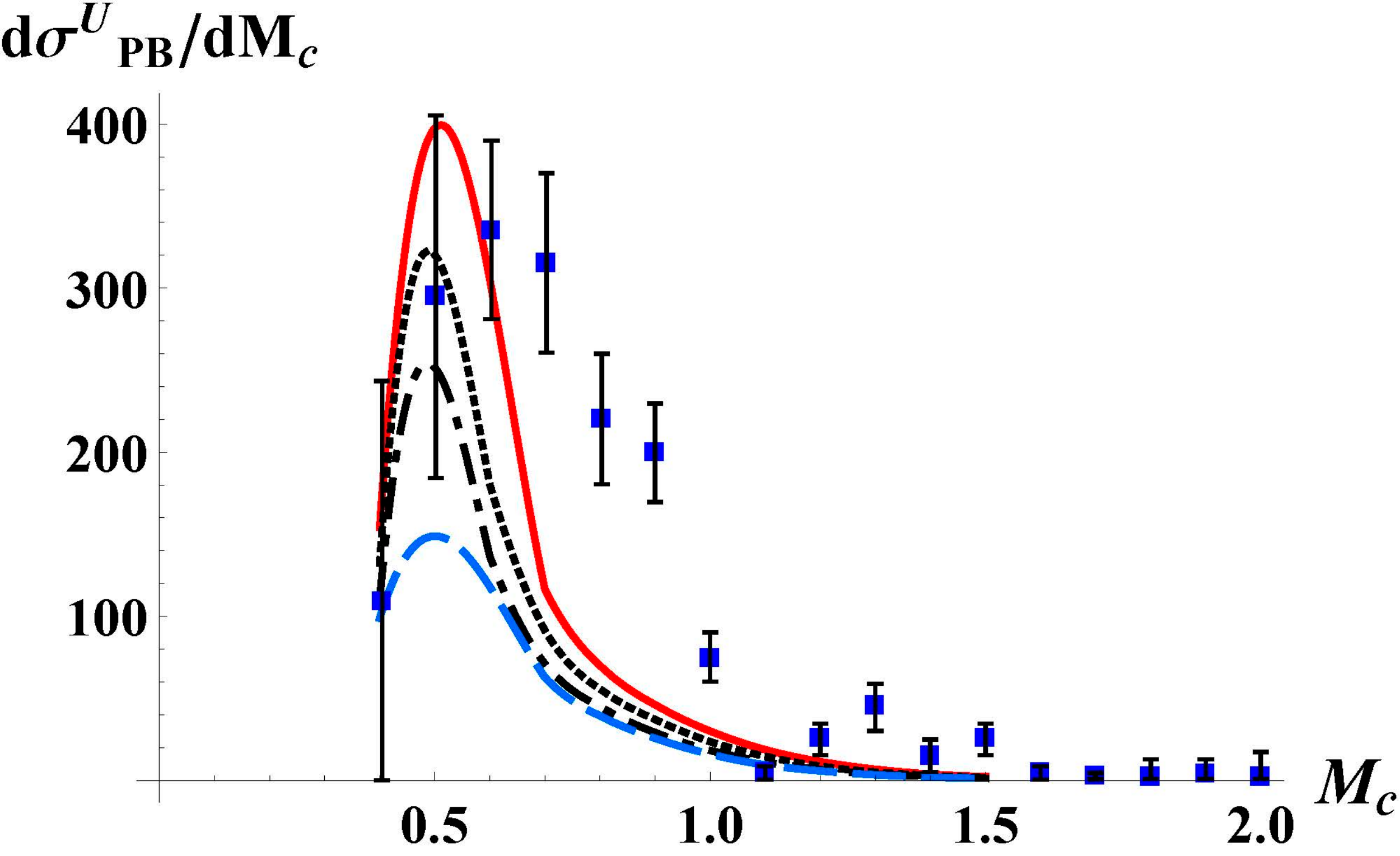}\\
c)\includegraphics[width=0.35\textwidth]{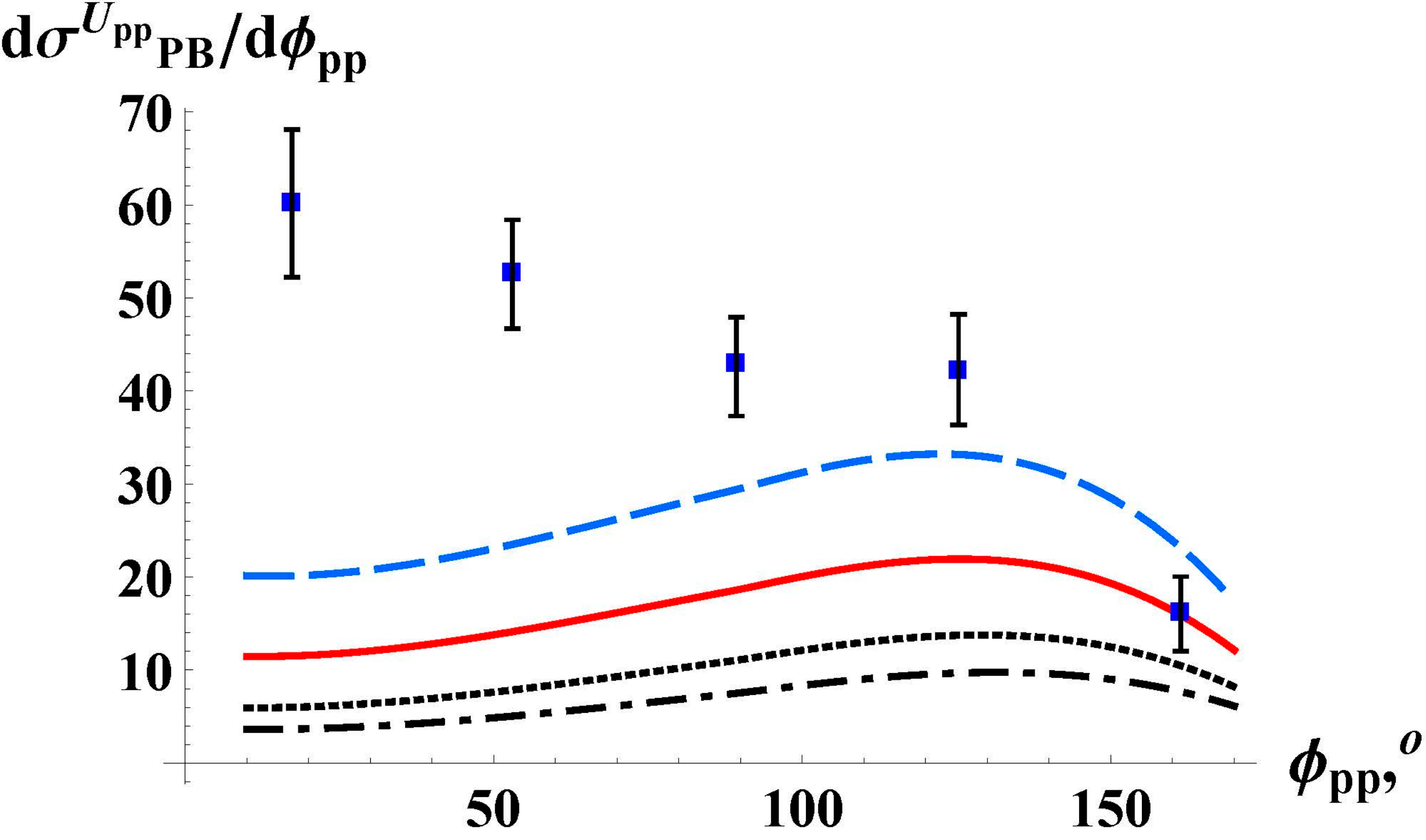}\\
d)\includegraphics[width=0.35\textwidth]{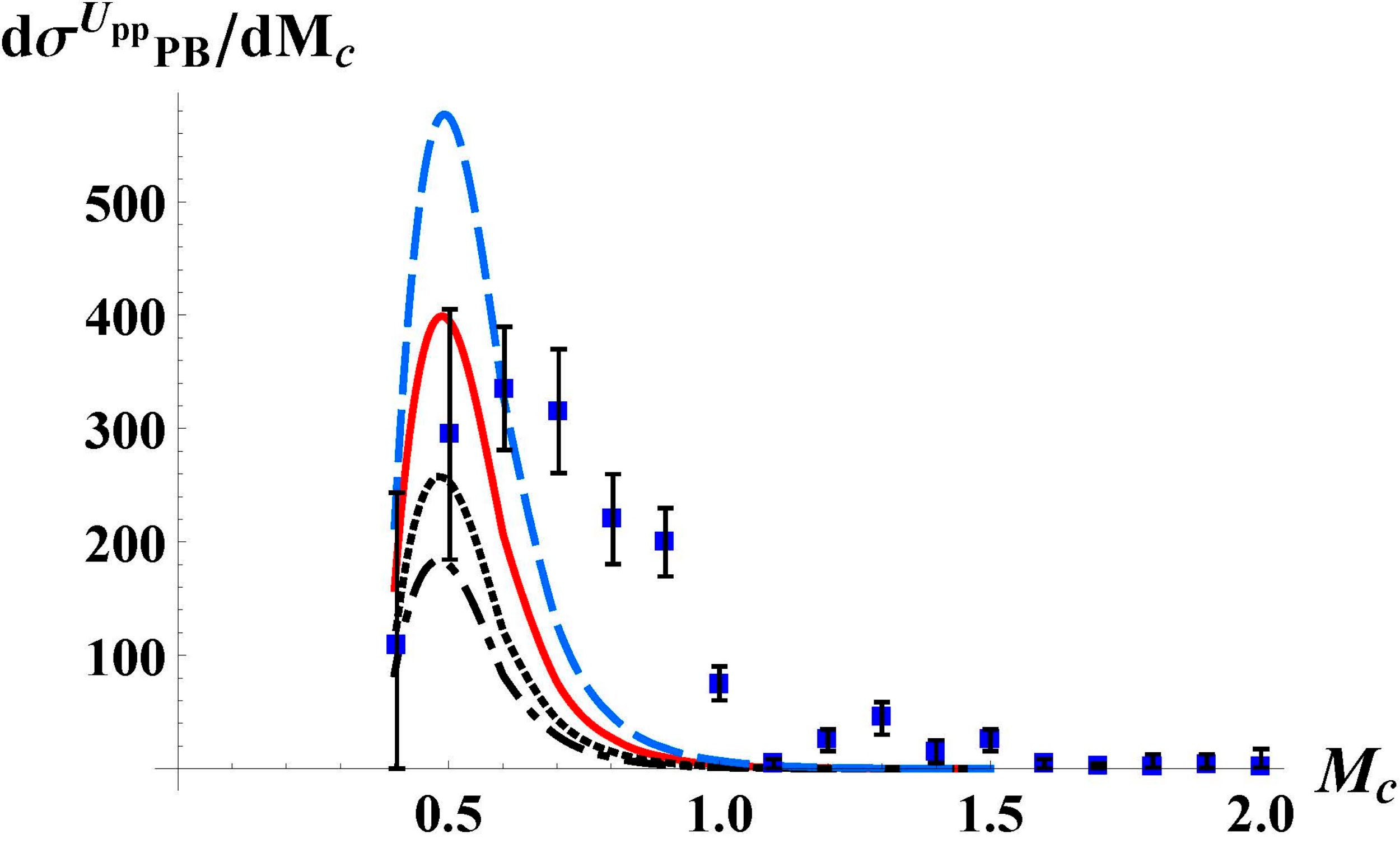}
\caption{\label{fig:starPB} PB case of the model (see Fig.~\ref{fig3:MY4CASES}). The data on the process $p+p\to p+\pi^++\pi^-+p$ 
at $\sqrt{s}=200$~GeV, $|\eta_{\pi}|<1$, $|\eta_{\pi\pi}|<2$, $p_{T\pi}>0.15$~GeV, $0.005<-t_{1,2}<0.03$~GeV$^2$, (STAR collaboration~\cite{STARdata1},\cite{STARdata2}).  {\bf a, b} The result 
with all p p and $\pi$ p rescattering corrections,  
{\bf c, d} show the result when we try to fit the data by formulas without $\pi$ p rescattering terms. Curves from 
up to down correspond to different values of the parameter $\Lambda_{\pi}$ in the off-shell 
pion form factor~(\ref{eq:offshellFpi}): {\bf a, b} $\Lambda_{\pi}=0.45,0.43,0.41,0.4$~GeV, {\bf c, d} $\Lambda_{\pi}=0.43,0.4,0.37,0.35$~GeV.
}
\end{center}
\end{figure}

\hspace*{0.5cm} In this section the data of the STAR \linebreak col\-la\-bo\-ra\-tion~\cite{STARdata1},\cite{STARdata2} and model curves for different cases of Fig.~\ref{fig3:MY4CASES}
are presented. In our approach we have only one free parameter, $\Lambda_{\pi}$, that is why all the distributions are depicted for its different 
values. Also in every case we consider two possibilities: fitting data by formulas with 
all rescattering corrections (two upper pictures) and also in the approach with proton--proton rescattering 
only (i.e. fitting the data by formulas without pion--proton interactions in the final state, two lower pictures). 
We change $\Lambda_{\pi}$ and try to get the best description. As you can see
from Figs.~\ref{fig:starRF}--\ref{fig:starPB}, the best description is given in the RF case for both possibilities (see Fig.~\ref{fig:starRF}). Since the final pion--proton interaction can give rather large
suppression (about 10--20\%, as in Fig.~\ref{fig:CMS7RFX}), in our further calculations we use the full amplitude as depicted in  Fig.~\ref{fig3:MY4CASES} for the RF case. The RF case without pion--proton interactions in the final state (with its own values of $\Lambda_{\pi}$ for the best data description) we will show just for cheking of this possibility. 

\subsection{ISR and CDF data versus RF case of the model}

\begin{figure}[h!]
\begin{center}
a)\includegraphics[width=0.35\textwidth]{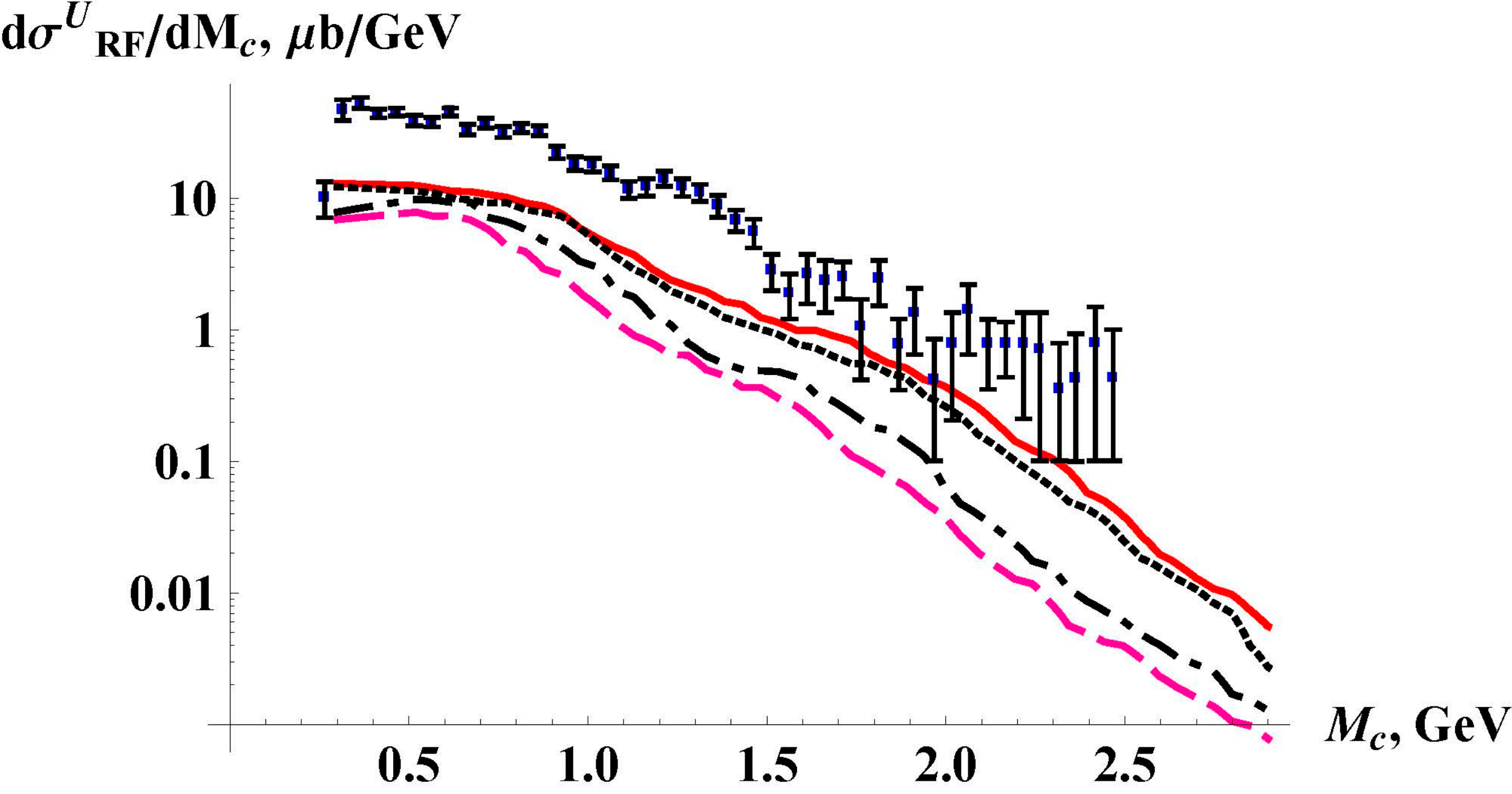}\\
b)\includegraphics[width=0.35\textwidth]{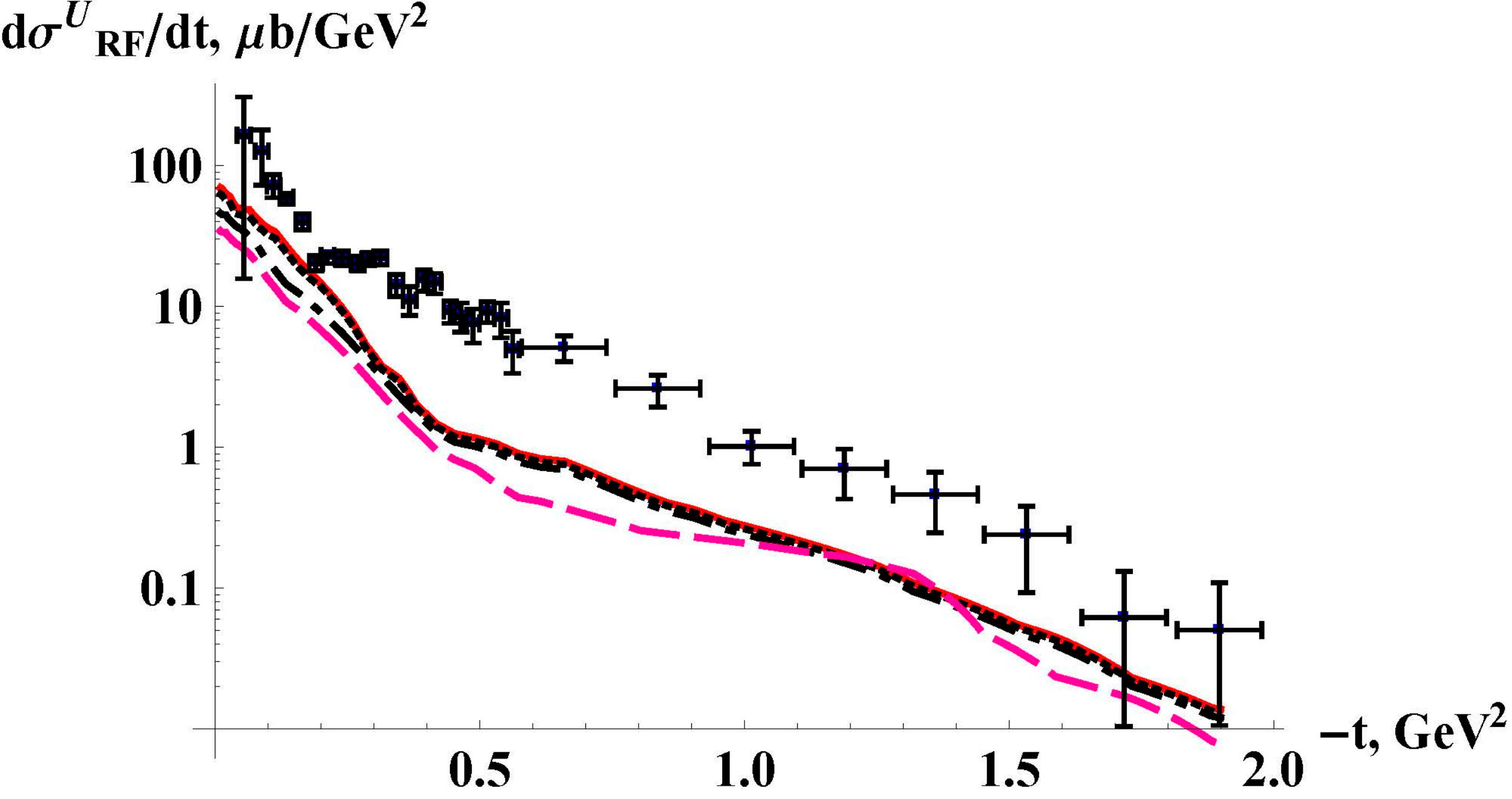}\\
c)\includegraphics[width=0.35\textwidth]{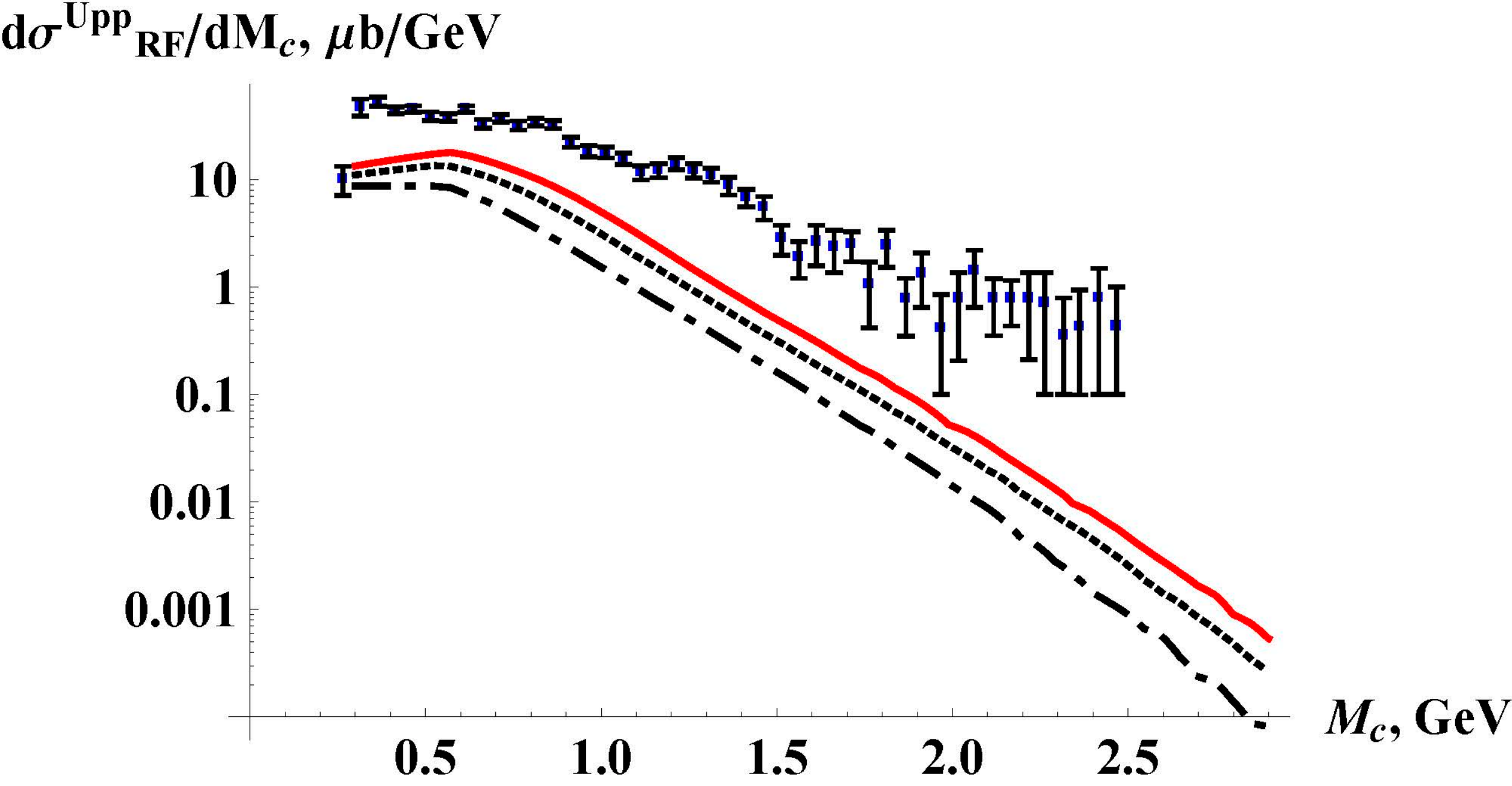}\\
d)\includegraphics[width=0.35\textwidth]{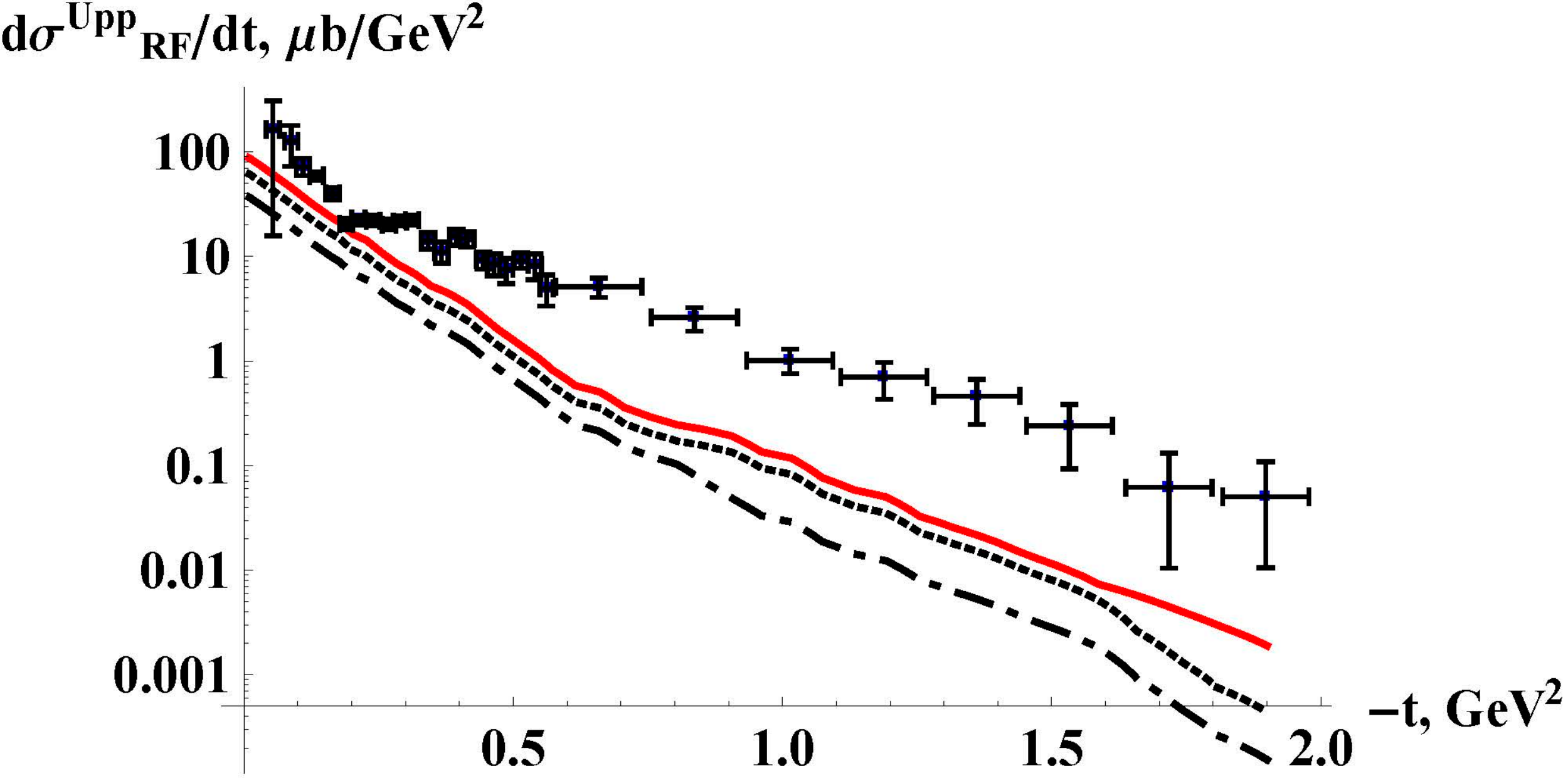}
\caption{\label{fig:ISR1RF} RF case of the model (see Fig.~\ref{fig3:MY4CASES}) with all rescattering corrections ({\bf a, b}) and also when we try to fit the data by formulas without final pion--proton 
rescattering ({\bf c, d}). The data on the process $p+p\to p+\pi^++\pi^-+p$ 
at $\sqrt{s}=63$~GeV, $|y_{\pi}|<1$, $\xi_p>0.9$, (ISR and ABCDHW collaborations~\cite{ISRdata1},\cite{ISRdata2}).  Curves from 
up to down correspond to different values of the parameter $\Lambda_{\pi}$ in the off-shell 
pion form factor~(\ref{eq:offshellFpi}): {\bf a, b} $\Lambda_{\pi}=4,3,1.6,1.2$~GeV, {\bf c, d} $\Lambda_{\pi}=1.2,1,0.8$~GeV.
}
\end{center}
\end{figure}

\begin{figure}[ht!]
\begin{center}
a)\includegraphics[width=0.35\textwidth]{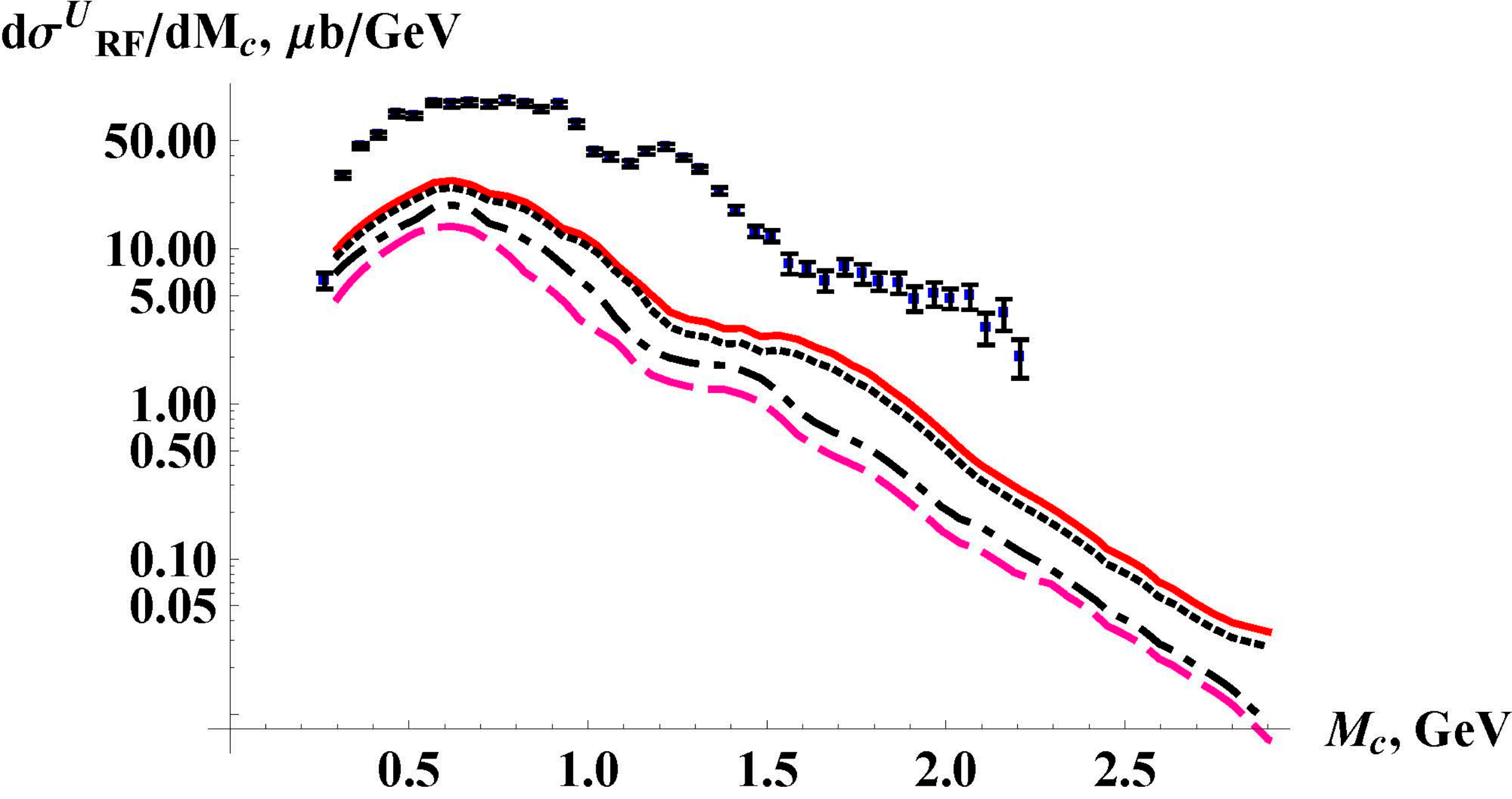}\\
b)\includegraphics[width=0.35\textwidth]{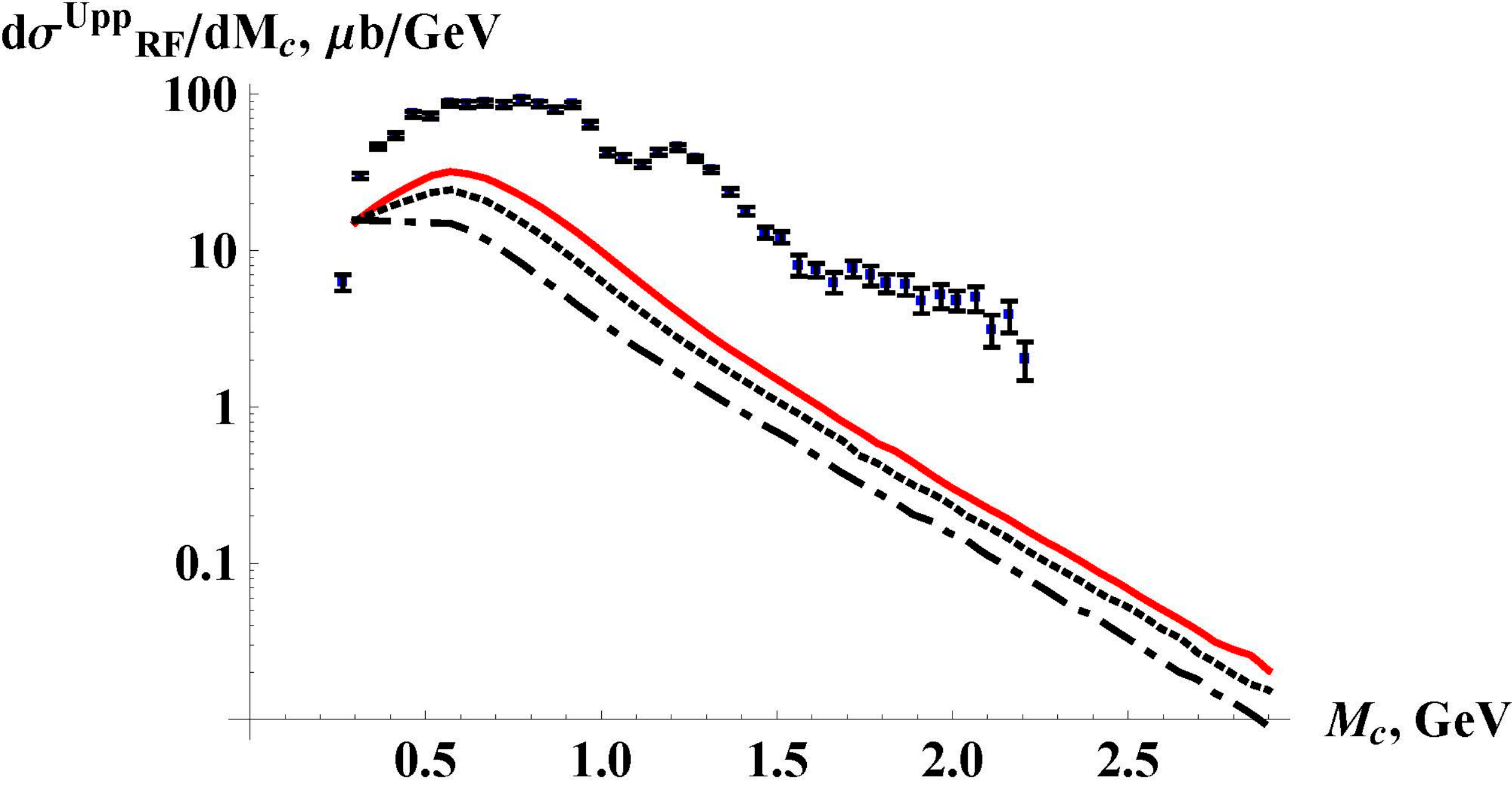}
\caption{\label{fig:ISR2RF} RF case of the model (see Fig.~\ref{fig3:MY4CASES}) with all rescattering corrections ({\bf a}) 
	and also when we try to fit the data by formulas without final pion--proton 
	rescattering  ({\bf b}). The data on the process $p+p\to p+\pi^++\pi^-+p$ 
at $\sqrt{s}=62$~GeV, $|y_{\pi}|<1.5$, $\xi_p>0.9$, (ISR and ABCDHW collaborations~\cite{ISRdata1},\cite{ISRdata2}).  Curves from 
up to down correspond to different values of the parameter $\Lambda_{\pi}$ in the off-shell 
pion form factor~(\ref{eq:offshellFpi}): {\bf a} $\Lambda_{\pi}=4,3,1.6,1.2$~GeV, {\bf b} $\Lambda_{\pi}=1.2,1,0.8$~GeV.
}
\end{center}
\end{figure}

\begin{figure}[hbt!]
\begin{center}
a)\includegraphics[width=0.35\textwidth]{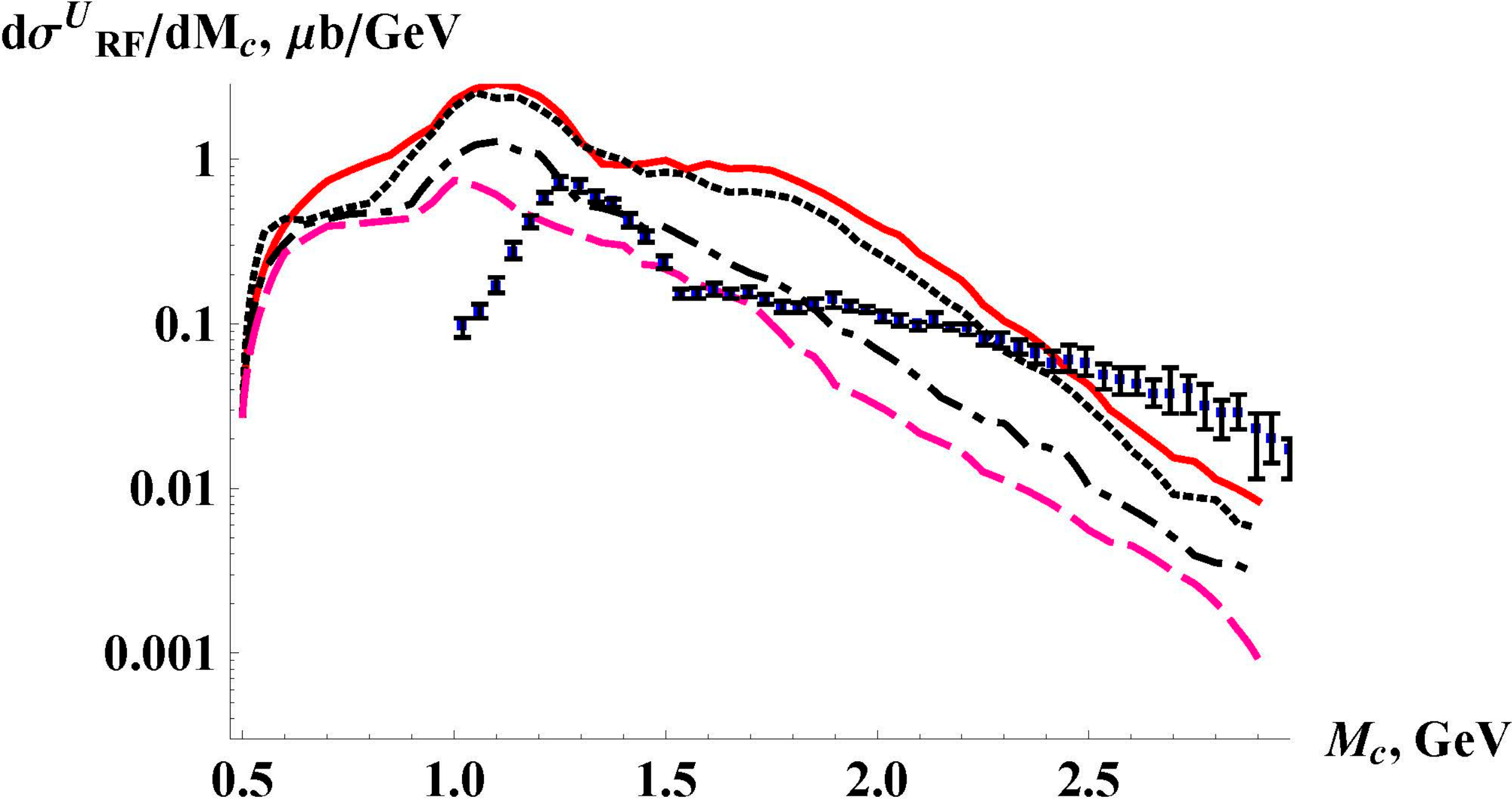}\\
b)\includegraphics[width=0.35\textwidth]{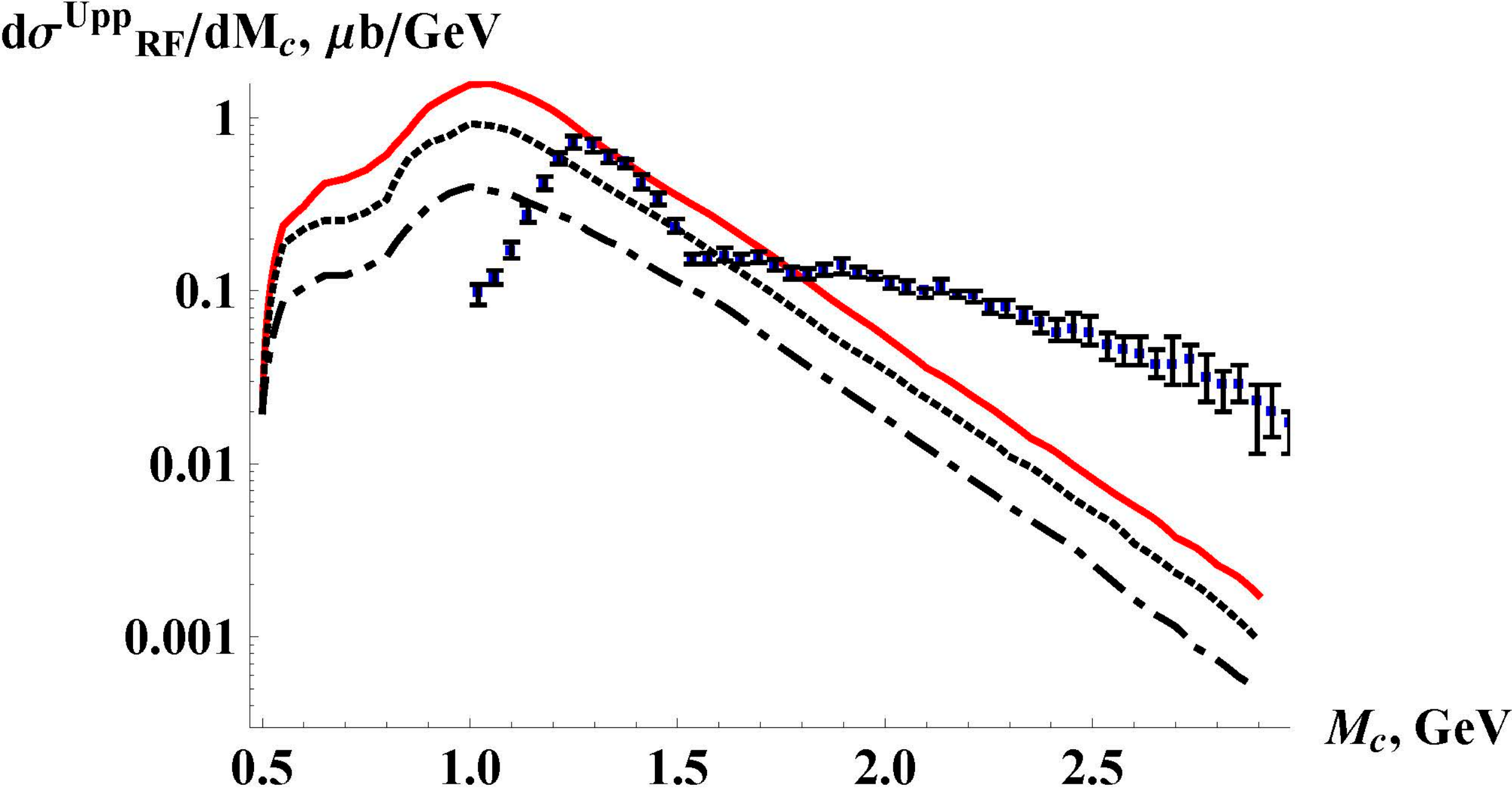}
\caption{\label{fig:CDF1RF} RF case of the model (see Fig.~\ref{fig3:MY4CASES}) with all rescattering corrections ({\bf a}) 
	and also when we try to fit the data by formulas without final pion--proton 
	rescattering  ({\bf b}). The data on the process $p+\bar{p}\to p+\pi^++\pi^-+\bar{p}$ 
at $\sqrt{s}=1.96$~TeV, $|\eta_{\pi}|<1.3$, $|y_{\pi\pi}|<1$, $p_{T,\pi}>0.4$~GeV, (CDF collaboration~\cite{CDFdata1},\cite{CDFdata2}).  Curves from 
up to down correspond to different values of the parameter $\Lambda_{\pi}$ in the off-shell 
pion form factor~(\ref{eq:offshellFpi}): {\bf a} $\Lambda_{\pi}=4,3,1.6,1.2$~GeV, {\bf b} $\Lambda_{\pi}=1.2,1,0.8$~GeV.
}
\end{center}
\end{figure}

\begin{figure}[hbt!]
\begin{center}
a)\includegraphics[width=0.35\textwidth]{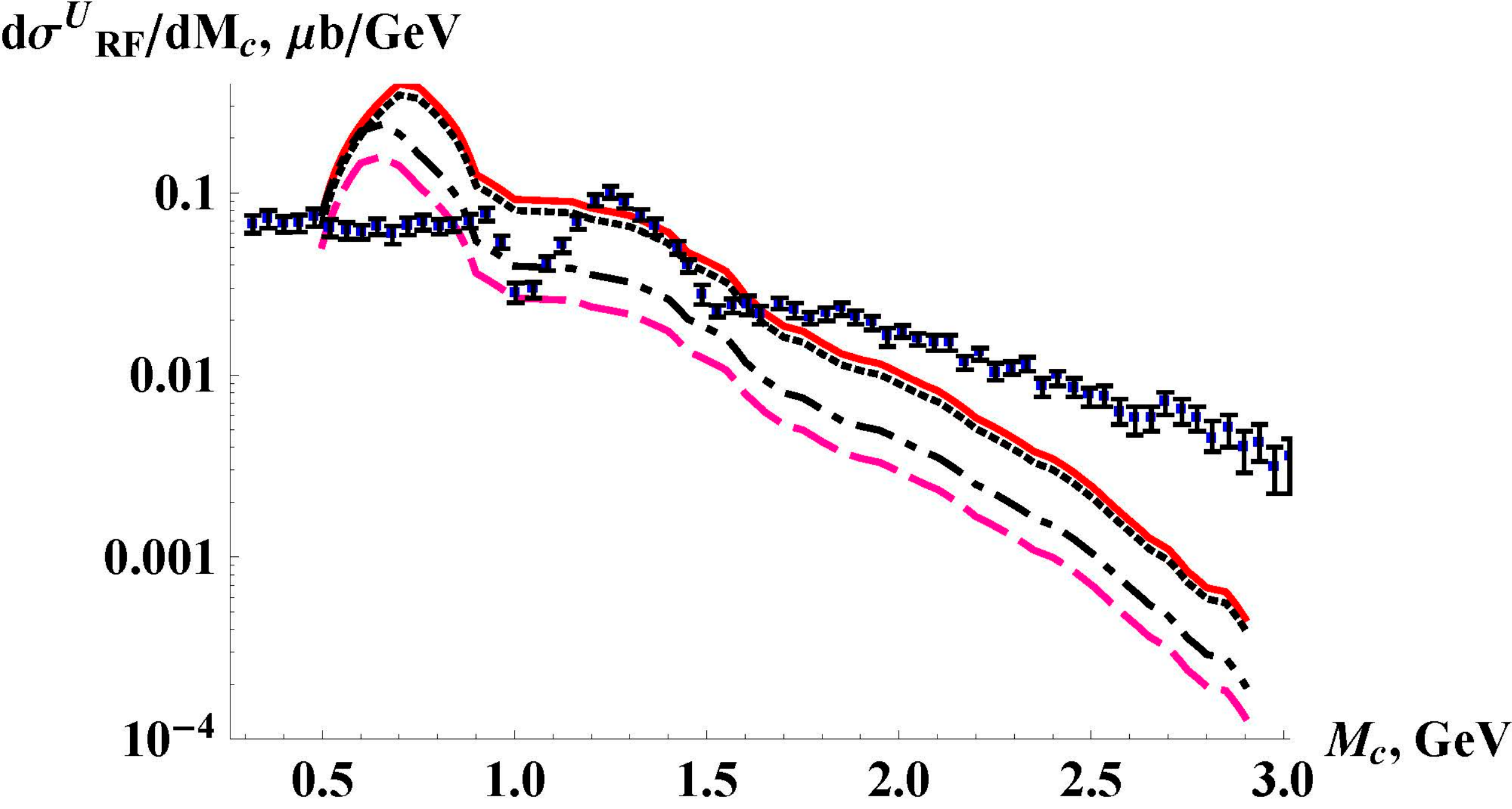}\\
b)\includegraphics[width=0.35\textwidth]{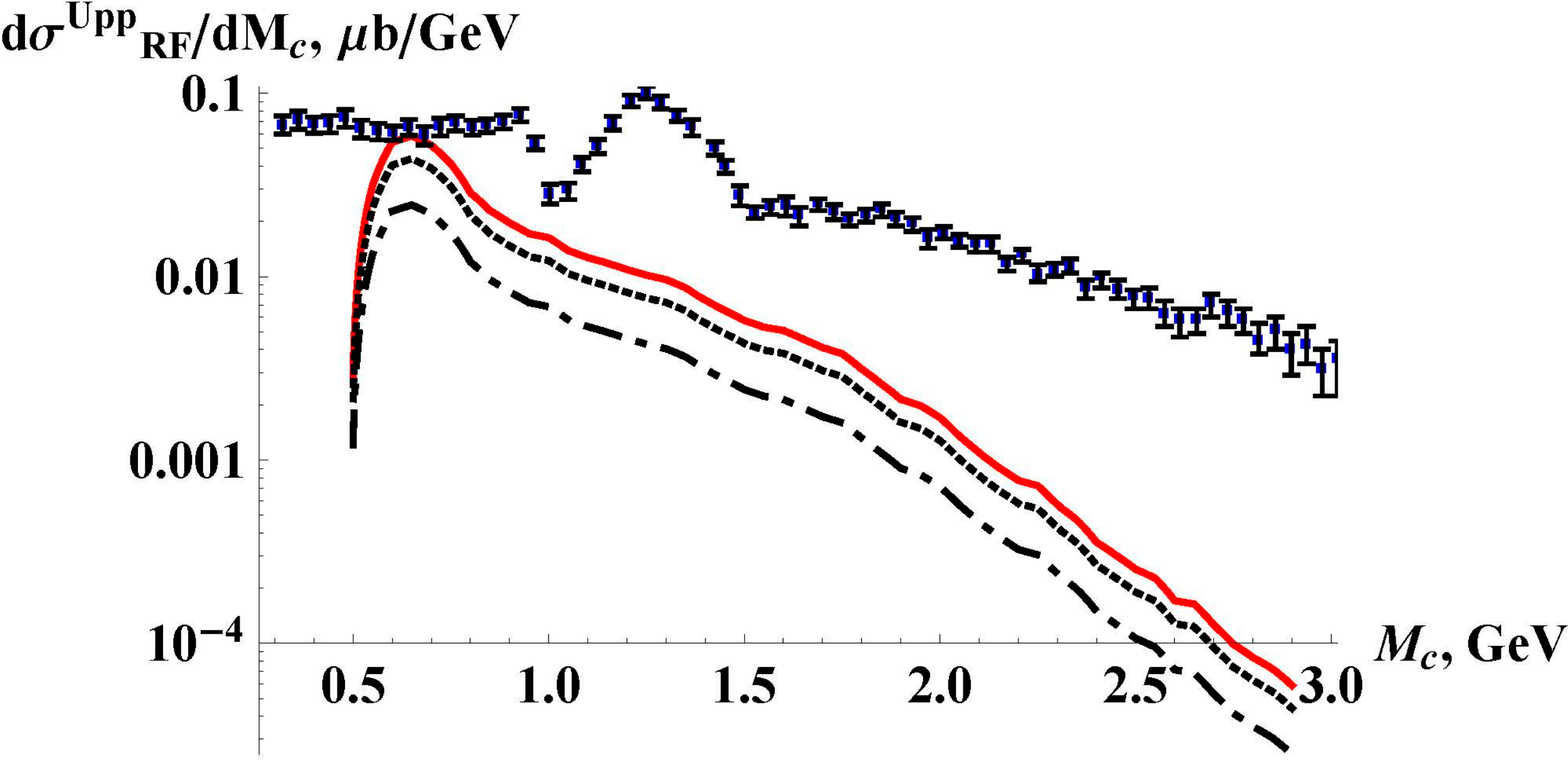}
\caption{\label{fig:CDF2RF} RF case of the model (see Fig.~\ref{fig3:MY4CASES}) with all rescattering corrections ({\bf a}) and also when we try to fit the data by formulas without final pion--proton 
	rescattering  ({\bf b}). The data on the process $p+\bar{p}\to p+\pi^++\pi^-+\bar{p}$ 
at $\sqrt{s}=1.96$~TeV, $|\eta_{\pi}|<1.3$, $|y_{\pi\pi}|<1$, $p_{T,\pi}>0.4$~GeV, $p_{T, \pi\pi}>1$~GeV, (CDF collaboration~\cite{CDFdata1},\cite{CDFdata2}).  Curves from 
up to down correspond to different values of the parameter $\Lambda_{\pi}$ in the off-shell 
pion form factor~(\ref{eq:offshellFpi}): {\bf a} $\Lambda_{\pi}=4,3,1.6,1.2$~GeV, {\bf b} $\Lambda_{\pi}=1.2,1,0.8$~GeV.
}
\end{center}
\end{figure}

Let us look at the ISR~\cite{ISRdata1},\cite{ISRdata2} and CDF~\cite{CDFdata1},\cite{CDFdata2} data with 
parameter $\Lambda_{\pi}$, which we use to describe the data from STAR
collaboration. Different cases are depicted on Figs.~\ref{fig:ISR1RF}-\ref{fig:CDF2RF}.

We see an underestimation of the ISR data. For these low energies 
we have to take into account possible corrections
to pion--proton amplitudes, since
our approach describes data well
only for energies greater than $\sim 3$~GeV. In each 
shoulder ($T_{\pi p}$ amplitude in Fig.~\ref{fig3:MY4CASES} RF) the energy can be less than $3$~GeV.

As to the CDF data (Figs.~\ref{fig:CDF1RF}, \ref{fig:CDF2RF}), which is overestimated for 
$M_{\pi\pi}<1.5$~GeV, we can
say that there are corrections (destructive interference terms) from resonances to the amplitude, like in 
Fig.~3 of~\cite{CEDPw4},\cite{CEDPw5} and other effects for low $M_{\pi\pi}$, for example,
the interference with $\gamma\gamma$ or $\gamma\mathbb{O}$ fusion in the central production process. Also
we have to take into account effects related to the irrelevance and possible modifications of the 
Regge approach (for the virtual pion exchange) in this kinematical region, as was discussed
in the introduction.

\begin{figure}[hbt!]
	\begin{center}
	\includegraphics[width=0.45\textwidth]{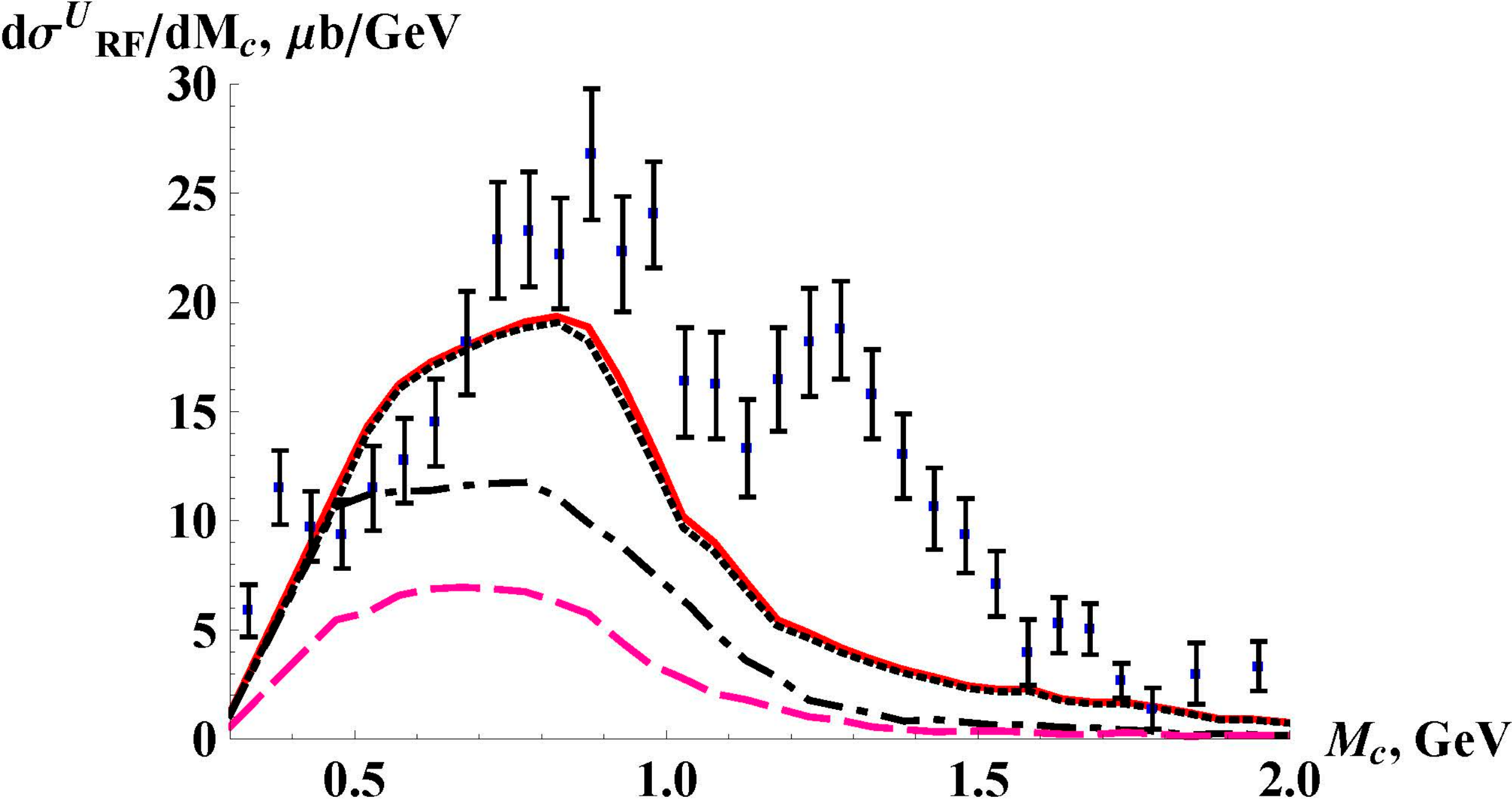}		
		\caption{\label{fig:CMS7RF} RF case of the model (see Fig.~\ref{fig3:MY4CASES}) with all rescattering corrections. The data on the process $p+\bar{p}\to p+\pi^++\pi^-+\bar{p}$ 
			at $\sqrt{s}=7$~TeV, $|y_{\pi}|<2$, $p_{T,\pi}>0.2$~GeV, (CMS collaboration~\cite{CMSdata1},\cite{CMSdata2}).  Curves from 
			up to down correspond to different values of the parameter $\Lambda_{\pi}$ in the off-shell 
			pion form factor~(\ref{eq:offshellFpi}):  $\Lambda_{\pi}=4,3,1.6,1.2$~GeV.
		}
	\end{center}
\end{figure}   

\begin{figure}[hbt!]
	\begin{center}
		\includegraphics[width=0.45\textwidth]{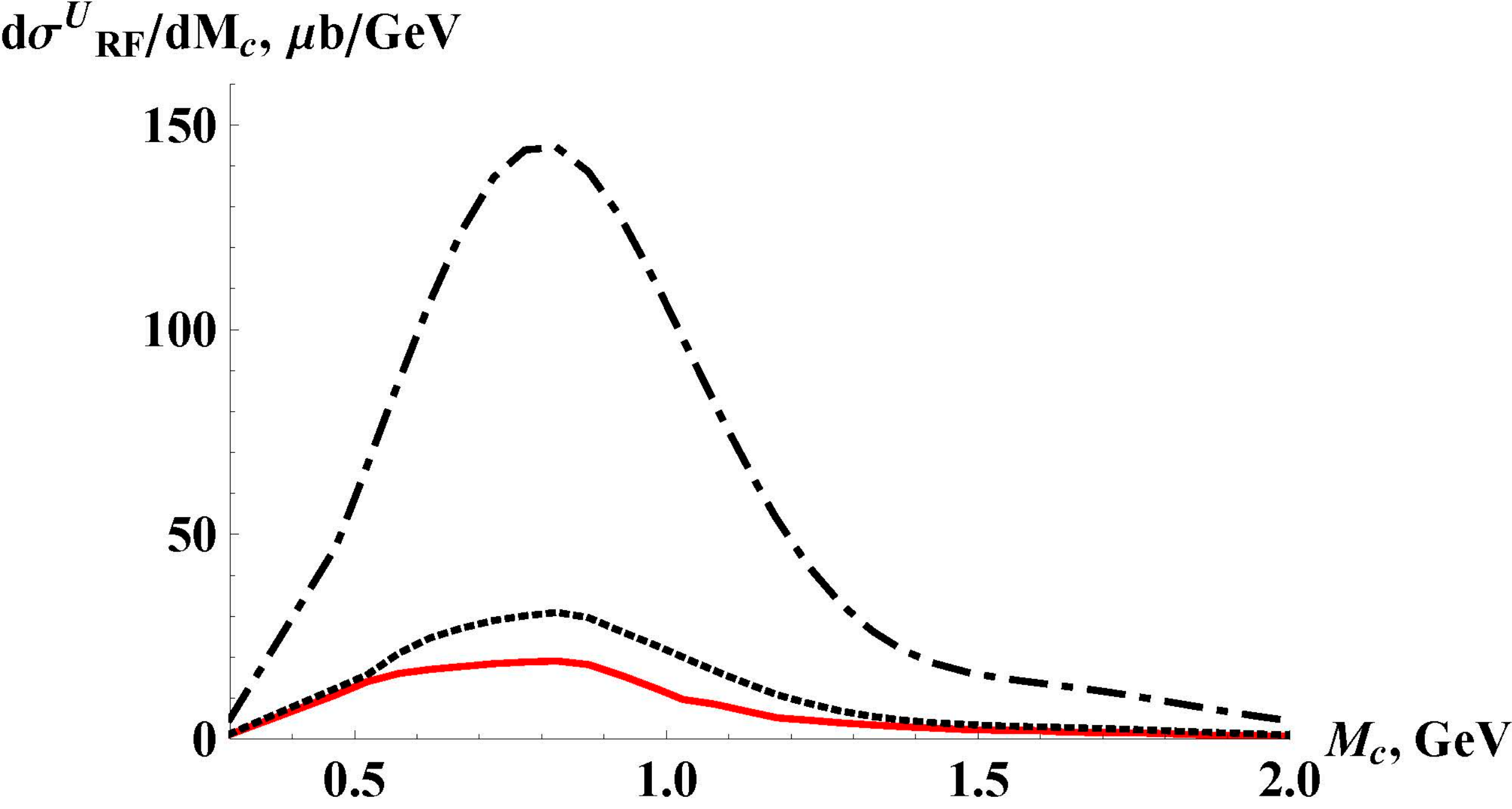}		
		\caption{\label{fig:CMS7RFX} RF case of the model (see Fig.~\ref{fig3:MY4CASES}) for the CMS energy.  Curves from 
			up to down correspond to the Born term, the amplitude with proton--proton rescattering corrections only and the one with 
			all the corrections (proton--proton and pion--proton). $\Lambda_{\pi}=3$~GeV.
		}
	\end{center}
\end{figure}   

\subsection{CMS data and predictions}

In Fig.~\ref{fig:CMS7RF} one can see the recent data from the CMS collaboration
and curves of our model. The upper curve, which corresponds to the parameter $\Lambda_{\pi}$, which better fits the STAR data on $\phi_{pp}$ distribution (but gives higher values for $M_{\pi\pi}>1$~GeV as depicted in Fig.~\ref{fig:starRF}b), also
describes the data of CMS collaboration well (but overestimates the CDF data, as was shown in 
Figs.~\ref{fig:CDF1RF}, \ref{fig:CDF2RF}). The lower curve 
underestimates the data 
from STAR, ISR and CMS, but it is close to the 
CDF data. Interference with resonances and modifications of the model in the region $|\hat{t}|\sim\hat{s}/2\sim 1$~GeV$^2$ 
can change the picture, especially for low $M_{\pi\pi}$, that is why
we have to take it into account when fitting the data (Fig.~\ref{fig:CMS7RFX}).

\subsection{Summary}

After the experimental data analysis we have several facts:
\begin{itemize}
\item In our approach the best description is given by the case RF (Fig.~\ref{fig3:MY4CASES}). That
is why effects from rescattering (unitarity) corrections are very important. 
\item The result is crucially dependent on the choice of $\Lambda_{\pi}$ in the off-shell pion form factor, i.e.
on $\hat{t}$ (virtuality of the pion) dependence. 
\item If we try to fit the data from STAR~\cite{STARdata1},\cite{STARdata2}, we find that the best description
gives overestimation of the CDF data~\cite{CDFdata1},\cite{CDFdata2} (especially in the region $M_{\pi\pi}< 1.5$~GeV)
and underestimation of the ISR data~\cite{ISRdata1},\cite{ISRdata2}. This is due to effects
like the interference with resonance contributions or $\gamma\gamma\to\pi\pi$ and $\gamma\mathbb{O}\to\pi\pi$ processes, 
effects related to the irrelevance and possible modifications of the 
Regge approach (for the virtual pion exchange) in this kinematical region, as was discussed
in the introduction, corrections to 
pion--pion scattering at low $M_{\pi\pi}$, and corrections to
$T_{\pi p}(s,t)$ for $\sqrt{s}<3$~GeV; 
\item Predictions for CMS are close to the data, if we use the best fit to the STAR data on $\phi_{pp}$ distribution (see Fig.~\ref{fig:starRF}b). We need also an 
estimate of the interference with resonance terms to see the full picture and draw  final conclusions.  
\end{itemize}

\section{Pomeron--Pomeron to pion--pion cross-section}

Another interesting question, which we can discuss here, concerns the Pomeron--Pomeron 
cross-section. As was shown in~\cite{mySD}, it is possible to extract the Pomeron--proton cross-section 
from the data on single (SD) and
double (DD) dissociation, and this numerical value appears to be of the order of typical hadron--hadron cross-sections. It was
done by the use of covariant reggeization method with conserved spin-J meson currents,\linebreak which helps to solve
the old problem of the very small Po\-me\-ron--pro\-ton cross-sec\-tion extracted by other\linebreak au\-thors~\cite{kaidalov1}-\cite{kaidalov2}. 

Reggeon--hadron and reggeon--reggeon scattering can be considered as a scattering of all possible real mesons lying on the Regge trajectory 
of hadrons. Conceptually it is similar to hydrogen--hadron or hydrogen--hydrogen scattering, since hydrogen has the spectrum of the states, and each of them has its own probability to scatter on a hadron or another hydrogen atom. A specific feature is
that we deal in this case with ``off-shell atoms''. There is some misunderstanding concerning the physical 
nature of reggeons. Actually, the reggeon is quite a general notion to qualitatively describe a 
generic quantum composite system. Conceptually a hadronic reggeon differs from the familiar hydrogen atom only by constituent content: electron--proton in the latter case and, say, quark--antiquark in the former. Energy levels of 
both are given by the corresponding Regge trajectories. So the reggeon is a 
full fledged composite particle in 
the same sense as an atom. Certainly the specific properties are 
different due to different binding forces.

As was shown also in~\cite{mySD}, absorptive corrections play the crucial role in high energy scattering, and
make the extraction procedure rather complicated and model dependent. We should propose
some appropriate pa\-ra\-met\-rization for Pomeron--hadron cross-section, then we apply unitarization procedure to
obtain real SD or DD cross-sections.

It is possible to perform a similar procedure to extract the Pomeron--Po\-me\-ron cross-section. In further work we shall consider
the extraction of the total Pomeron--Po\-me\-ron cross-section. Here we restrict ourselves by the extraction
of the Pomeron--Pomeron to pion--pion one. Let us use the parametrization~(\ref{eq:MU}) for CEDP di-pion production to
fix one parameter $\Lambda_{\pi}$ from the experimental data. After that we can simply estimate the 
Pomeron--Pomeron to di-pion cross-section (we use the covariant method from Appendix~C):
\begin{eqnarray}
&&\!\!\! \frac{\mathrm{d}\sigma_{J_1J_2\to\pi\pi}(\hat{s})}{\mathrm{d}\hat{t}} = 
\nonumber\\
&& \!\!\!
\frac{1}{(2J_1+1)(2J_2+1)}\frac{1}{16\pi\lambda(\hat{s},t_1,t_2)}
\left| {\cal P}_{\pi}(\hat{s},\hat{t}) \hat{F}_{\pi}(\hat{t})^2 \right|\times\nonumber\\
&& \!\!\!\sum_{\lambda_1,\lambda_2}\left| e_{\mu_1\ldots\mu_{J_1}}^{(\lambda_1)}(t_1)
e_{\nu_1\ldots\nu_{J_2}}^{(\lambda_2)}(t_2) 
F^{J_1J_2\to\pi\pi}_{\mu_1\ldots\mu_{J_1},\;\nu_1\ldots\nu_{J_2}} \right|^2=\nonumber
\end{eqnarray}

\begin{eqnarray}
&& \!\!\!
\frac{1}{(2J_1+1)(2J_2+1)}\frac{1}{16\pi\lambda(\hat{s},t_1,t_2)}
\left| {\cal P}_{\pi}(\hat{s},\hat{t}) \hat{F}_{\pi}(\hat{t})^2 \right|\times\nonumber\\
&& \!\!\!
\Pi_{\mu_1\ldots\mu_{J_1}}^{\mu^{\prime}_1\ldots\mu^{\prime}_{J_1}}(t_1)
\Pi_{\nu_1\ldots\nu_{J_2}}^{\nu^{\prime}_1\ldots\nu^{\prime}_{J_2}}(t_2)
W^{\mu_1\ldots\mu_{J_1},\;\nu_1\ldots\nu_{J_2}}_{\mu^{\prime}_1\ldots\mu^{\prime}_{J_1},\;\nu^{\prime}_1\ldots\nu^{\prime}_{J_2}} =\nonumber\\
&& \!\!\!\!\!
\frac{\left| {\cal P}_{\pi}(\hat{s},\hat{t}) \hat{F}_{\pi}(\hat{t})^2 g_{J_1}^{\pi}(t_1)g_{J_2}^{\pi}(t_2)
\right|^2\!\! \prod_{i=1}^2\frac{2^{J_i-1}(J_i-1)!J_i!}{(2J_i-1)!}}{(2J_1+1)(2J_2+1)\cdot 16\pi\lambda(\hat{s},t_1,t_2)}
\times\nonumber\\
&&  \times{\cal F}_{J_1}(t_1,\hat{t})  {\cal F}_{J_2}(t_2,\hat{t}).
\label{eq:JJtopipi}
\end{eqnarray}
Here 
$
\Pi^{\mu^{\prime}_1\ldots}_{\mu_1\ldots}=\sum_{\lambda} e^{(\lambda)}_{\mu^{\prime}_1\ldots} e^{*(\lambda)}_{\mu_1\ldots}
$
is the structure in the propagator like~(\ref{eq:Propnonconserv}), $e^{(\lambda)}_{\mu_1\ldots}$ are polarization 
tensors,\linebreak
$F^{J_1J_2\to\pi\pi}_{\mu_1\ldots}$ is the Pomeron--Pomeron 
amplitude and 
$W^{\mu_1...}_{\mu^{\prime}_1\ldots}$ is the hadro\-nic tensor for this process 
made of Pomeron--pion--pion vertices $T_{\mu_1\ldots}(p,\Delta)$, where $g^{\pi}_J(t)$\linebreak $=F_J(t)/(m_{\pi}^2-t/4)^{J/2}$, $F_J$ is the leading form factor 
from (\ref{eq:Ttensor0}). After reggeization we get
\begin{eqnarray}
&& \frac{\mathrm{d}\sigma_{{\mathbb P}{\mathbb P}\to\pi\pi}(\hat{s})}{\mathrm{d}\hat{t}} = 
\prod_{i=1}^2\frac{2^{\alpha_{\mathbb P}(t_i)-1}\Gamma(\alpha_{\mathbb P}(t_i))\Gamma(\alpha_{\mathbb P}(t_i)+1)}{\Gamma(2\alpha_{\mathbb P}(t_i))}
\nonumber\\
&& \frac{\left| {\cal P}_{\pi}(\hat{s},\hat{t}) \hat{F}_{\pi}(\hat{t})^2 g^{\pi}(t_1)g^{\pi}(t_2)
	\right|^2 	 {\cal F}(t_1,\hat{t})  {\cal F}(t_2,\hat{t})
}{(2\alpha_{\mathbb P}(t_1)+1)(2\alpha_{\mathbb P}(t_2)+1)\cdot 16\pi\lambda(\hat{s},t_1,t_2)}
\label{eq:PPtopipi},\\
&& g^{\pi}(t)=\frac{\beta_{\mathbb{P}}(t)2^{\alpha_{\mathbb P}(t)}}{\pi\alpha^{\prime}_{\mathbb P}(t) g^p(t)}
\label{eq:gpi},
\end{eqnarray}
where all the functions are defined in Appendix~B. If we use the  approach~(\ref{eq:gpi}), where all terms like 
$$
2\sqrt{-t}\lambda^{1/2}(m_i^2,m_j^2,t)
$$ 
in~(\ref{UneqMass}) are absorbed into the residue, we have to multiply the result by the additional factor ${\cal F}(t_1,\hat{t}){\cal F}(t_2,\hat{t})$, where
\begin{equation}
{\cal F}(t,\hat{t})=
\left( 
\frac{m_{\pi}^2-\frac{\left(m_{\pi}^2-\hat{t}+t\right)^2}{4t}}{s_0}
\right)^{\alpha_{\mathbb P}(t)}
\label{eq:factorcons}
\end{equation}
for conserved currents,
and
\begin{equation}
{\cal F}(t,\hat{t})\simeq
\left( 
\frac{m_{\pi}^2-\frac{\left(m_{\pi}^2-\hat{t}+t\right)^2}{4m_J^2}}{s_0}
\right)^{\alpha_{\mathbb P}(t)}
\label{eq:factornoncons}
\end{equation}
is the leading term for the case of non-conserved currents (see~(\ref{eq:Propnonconserv})).
 
\begin{figure}[hbt!]
\begin{center}
a)\includegraphics[width=0.35\textwidth]{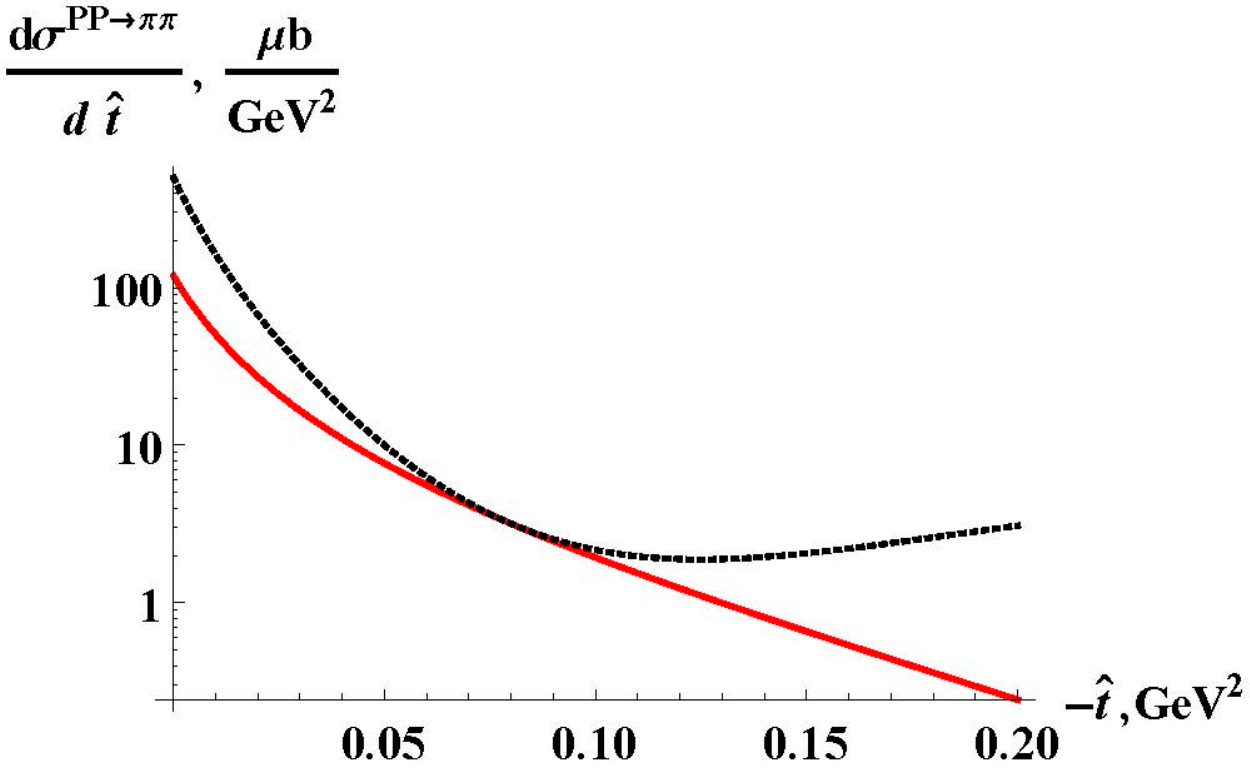}\\
b)\includegraphics[width=0.35\textwidth]{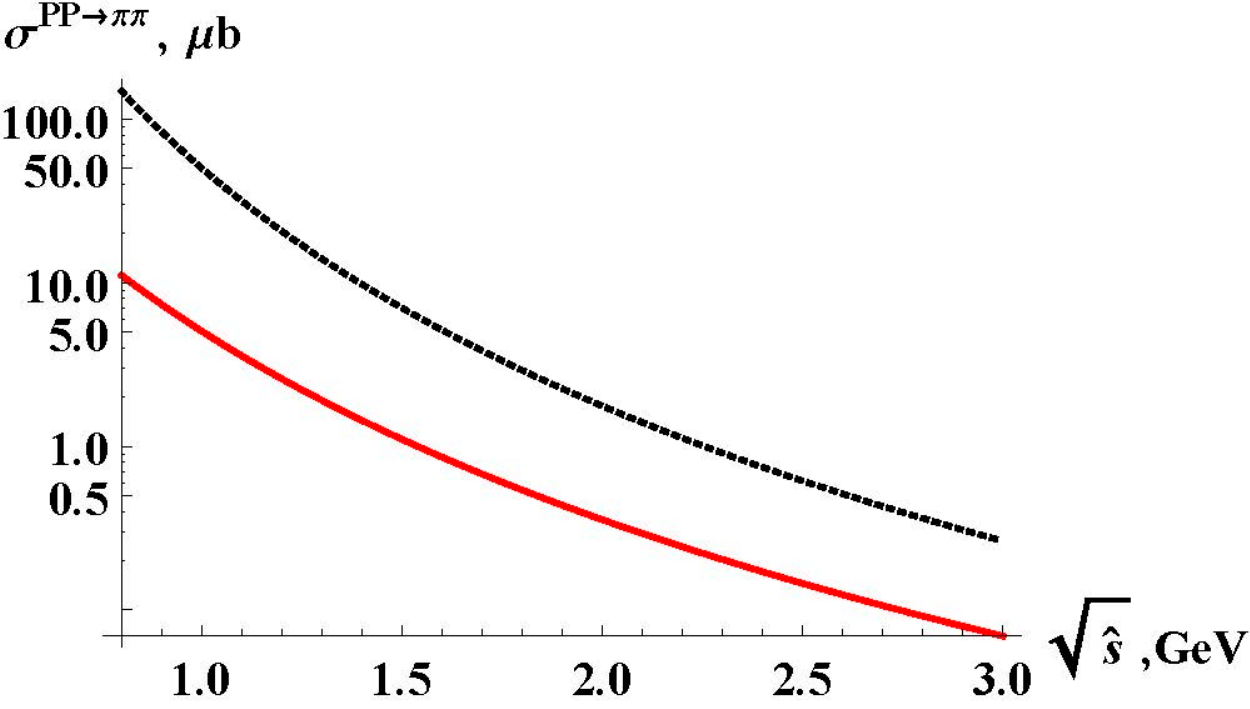}\\
c)\includegraphics[width=0.35\textwidth]{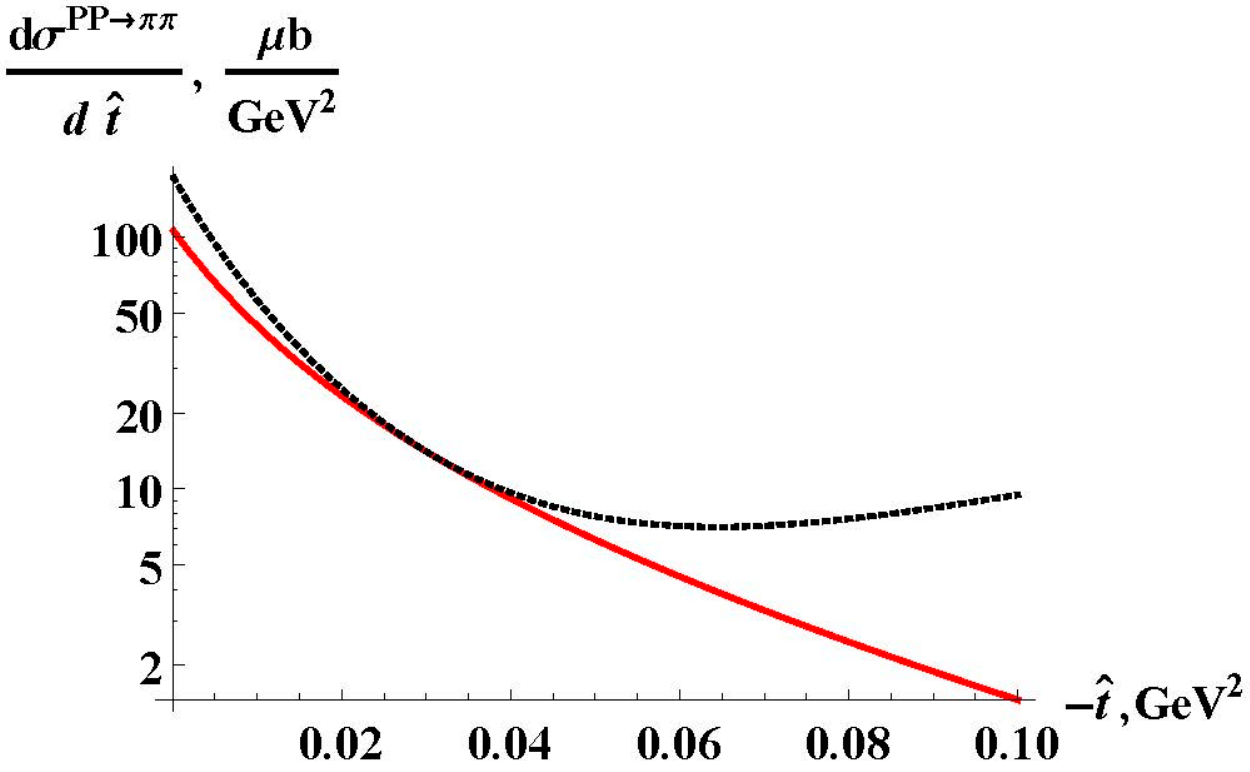}\\
d)\includegraphics[width=0.35\textwidth]{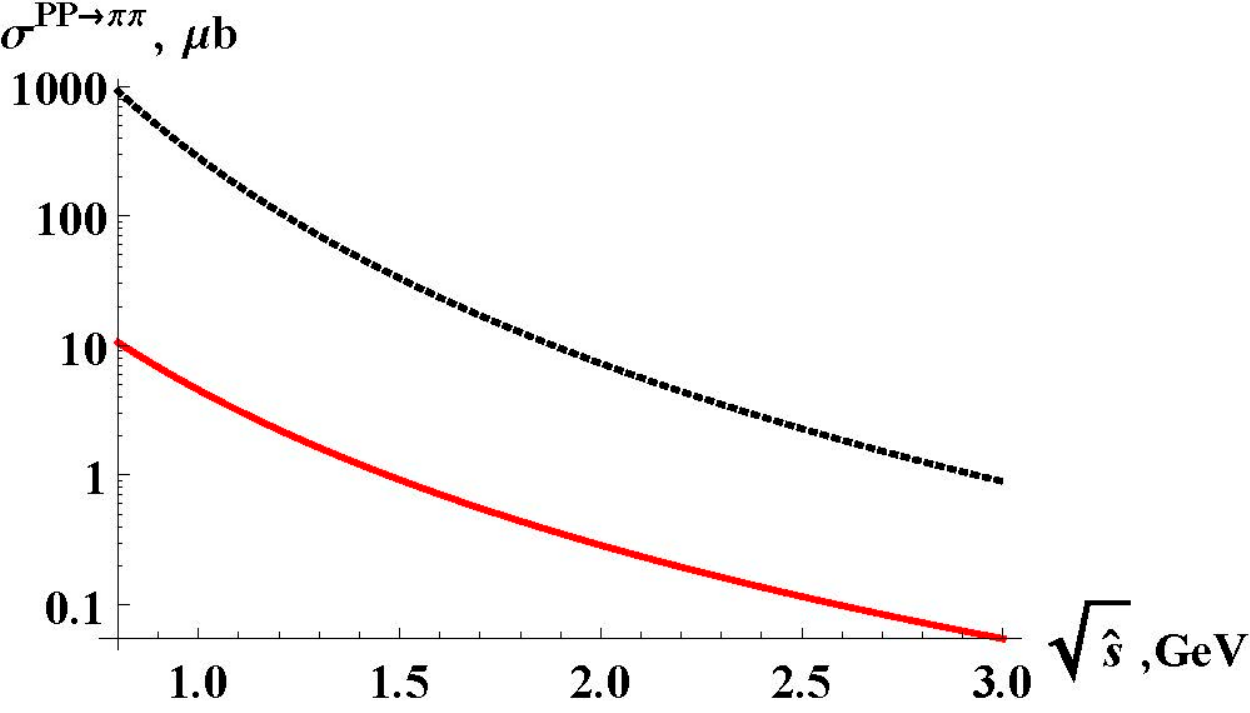}
\caption{\label{fig:PPtopipics} Results of calculations~(\ref{eq:PPtopipi})-(\ref{eq:factorcons}) for Pomeron--Pomeron to pion--pion cross-section 
for $t_{1,2}=-0.1$~GeV$^2$ (a,b) and  $t_{1,2}=-0.05$~GeV$^2$ (c,d). Two curves present cases of non-conserved (solid) and conserved (dotted)
hadronic currents. $\sqrt{\hat{s}}=1.5$~GeV in (a), (c). The parameter of the off-shell form factor is taken $\Lambda_{\pi}=1$~GeV.
}
\end{center}
\end{figure}   

Results of calculations~(\ref{eq:PPtopipi})--(\ref{eq:factornoncons}) are shown in Fig.~\ref{fig:PPtopipics}. In the case of 
conserved meson currents we obtain Po\-me\-ron--Po\-me\-ron to pion--pion cross-section $10\div 100$ times higher than in the case of non-conserved
currents and with more specific and strong dependence on the Pomeron virtuality. In the old work~\cite{oldPPtot1}-\cite{oldPPtot3} the extracted 
Po\-me\-ron--Po\-me\-ron total cross-section was of the order 100~$\mu \mathrm{b}$ at $\sqrt{\hat{s}}<3$~GeV and almost independent 
on Pomeron's virtuality. $\sigma_{\mathbb{P}\mathbb{P}\to\pi\pi}$ should be at least less than this number.\footnote{Let us note that 
	authors of~\cite{oldPPtot1}-\cite{oldPPtot3} extracted
	$\sigma^{tot}_{\mathbb{P}\mathbb{P}}$ in the classical approach which corresponds to the case of non-conserved currents in the 
	present paper, that is why we should compare their result with the solid curve on Fig.~\ref{fig:PPtopipics}b, which gives $\sigma_{\mathbb{P}\mathbb{P}\to\pi\pi}\sim 0.1\div 5$~$\mu \mathrm{b}$ for $\sqrt{s}<3$~GeV and $|\hat{t}|=0.1$~GeV$^2$ as found in~\cite{oldPPtot1}-\cite{oldPPtot3}. To compare our results for the case of conserved currents we need to use the same approach also
	when extract $\sigma^{tot}_{\mathbb{P}\mathbb{P}}$} Our
calculations in the same kinematical region give numbers of the order $0.1-5$~$\mu\mathrm{b}$ for non-conserved currents and 
$0.3-100$~$\mu \mathrm{b}$ for the case of conserved currents. There are strong contributions of 
other processes (especially production of resonances) in this region, that is why
we should have $\sigma_{\mathbb{P}\mathbb{P}\to\pi\pi}\ll\sigma^{tot}_{\mathbb{PP}}$, which is obvious in the case 
of non-conserved currents where we have both extracted numbers to compare.

In Sect.~\ref{sec:data} it is shown that RF and, possibly, PF modes (see Fig.~\ref{fig3:MY4CASES} for notations) 
give an appropriate description of the 
data, i. e. we have to take into account all rescattering corrections (even in the $T_{\pi p}$ 
amplitudes). If to use the RB mode, as some other authors do~\cite{CEDPw1}-\cite{CEDPw6},\footnote{Of course, they have fitted the data 
	on pion--proton elastic cross-sections, using the Born term only, but in our approach we use the full eikonalized amplitude. That is why their description of LM CEDP (in the RB mode) is also good, but only with their own parameters. In our approach the RB mode is not
	good in the data description.} it is possible
to extract the Pomeron--Po\-me\-ron cross-section more easily (``almost model independent method'', as was done, for example, for the 
pion--proton cross-section~\cite{myLHCf}).

\section*{Conclusions}

In this paper we have considered the process LM CEDP of di-pions and its description in the framework of the Regge-eikonal approach. 
Here we summarize all the facts and conclusions:
\begin{itemize}
\item After calculations of several cases (see Fig.~\ref{fig3:MY4CASES}) we can see that the 
RF case is better suited to describe the data on the process $p+p\to p+\pi^+ +\pi^- +p$.
\item When we try to fit the data from STAR collaboration~\cite{STARdata1},\cite{STARdata2} with
different values of $\Lambda_{\pi}$ (different behavior of the virtual pion form factor), we
obtain an underestimation of the ISR data~\cite{ISRdata1},\cite{ISRdata2} and an overestimation of 
the CDF~\cite{CDFdata1},\cite{CDFdata2}. Possible
explanations are: interference terms with 
resonances, corrections to $T_{\pi p}(s,t)$ for $\sqrt{s}<3$~GeV, off-shell 
pion effects, and some other mechanisms in Pomeron--Pomeron to pion--pion
process at low $M_{\pi\pi}$.
\item We have rather good predictions to the CMS 
data, when we use the fit to the STAR data on $\phi_{pp}$ distribution depicted on Fig.~\ref{fig:starRF}b, but
we have to take into account interference with resonances to see the full picture. These main 
open problems regarding the model parameters are related to interference terms (we have to know all couplings of pions to 
resonances), which require full spectrum simulation comparisons with data simultaneously in several differential 
observables. This should be done in further work.
\item After estimations of the Pomeron--Pomeron to pion--pion cross-section in the framework
of covariant reg\-geiza\-tion approach we obtain cross-sections which are (at least in the case of non-conserved 
Pomeron currents) much lower ($\sim 0.1\div 5\,\mu \mathrm{b}$) than
the total Pomeron--Pomeron cross-section, estimated by other authors~\cite{oldPPtot1}-\cite{oldPPtot3} ($\sim 100-300\,\mu \mathrm{b}$), which 
shows strong contributions from other processes (especially from resonance production). For the case of conserved currents
we need re-evaluate old results~\cite{oldPPtot1}-\cite{oldPPtot3} in the same approach.
\end{itemize}

In further work we will take into account possible modifications of the 
model (amplitudes for resonances in LM CEDP and their interference with di-pion one, pion--pion cross-section, additional 
off-shell effects in subamplitudes and so on) 
for the best description of the data. This model will be implemented 
to the Monte-Carlo event generator ExDiff~\cite{ExDiffmanual}. It is possible to 
calculate LM CEDP for other di-hadron final 
states ($p\bar{p}$ for ``Odderon'' hunting, $K^+K^-$, $\eta\eta^{\prime}$ and 
so on), which are also very informative for our understanding of diffractive mechanisms in strong interactions.


\section*{Acknowledgements}

I am grateful to Vladimir Petrov  and Anton Godizov for useful discussions.

\section*{Appendix A. Kinematics of LM CEDP}
\label{app:kinematics}

\begin{figure}[hbt!]
	\begin{center}
		\includegraphics[width=0.45\textwidth]{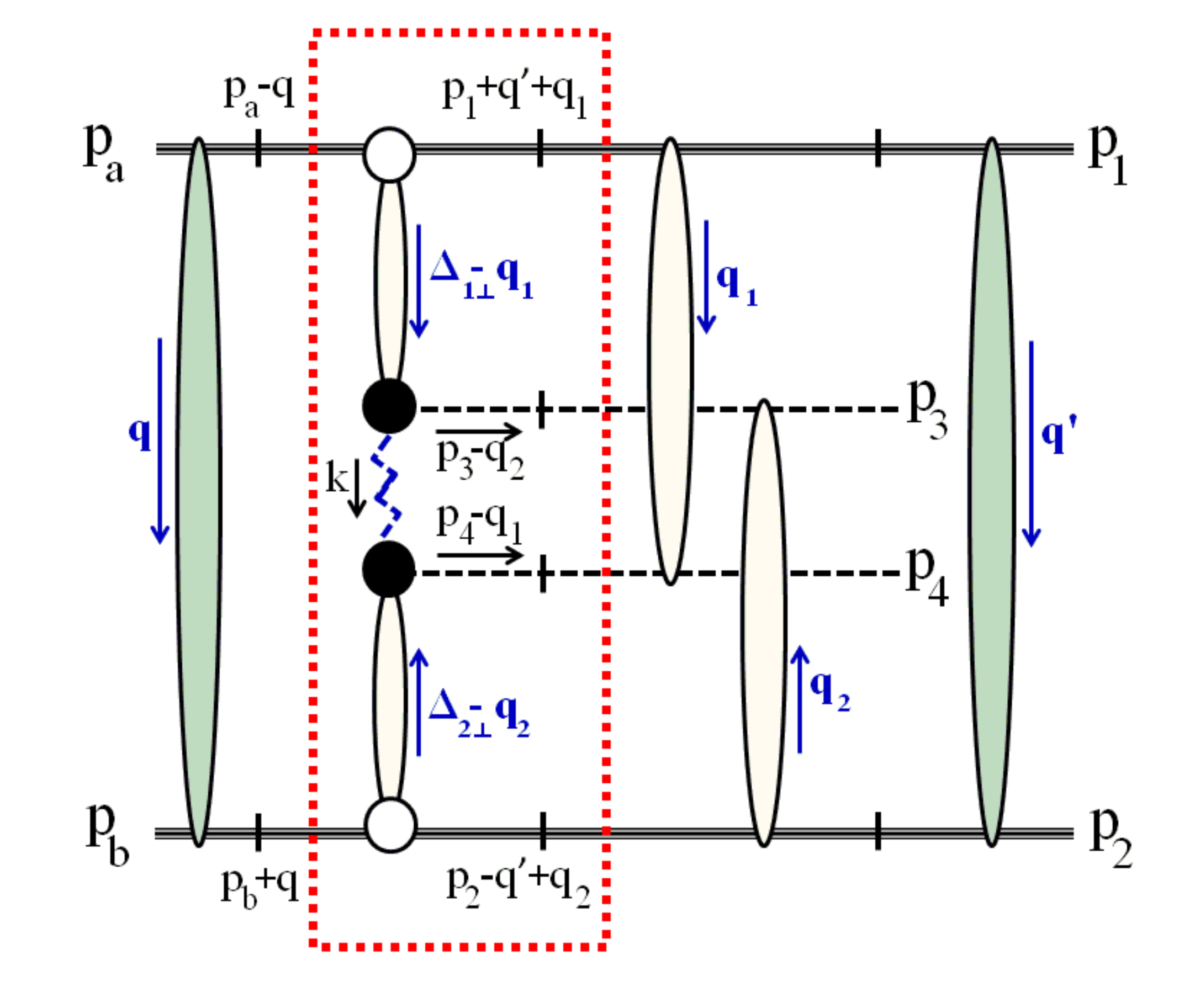}
		\caption{\label{fig4:totalkin} Total amplitude of the process of double pion LM CEDP $\left(p+p\rightarrow p+\pi^++\pi^-+p\right)$ with detailed kinematics. Proton--proton rescatterings in the initial and final states are depicted as black blobs, and pion--proton subamplitudes are shown as shaded blobes. All momenta are shown. The basic part of the amplitude, $M_0$ (see Eq.~(\ref{eq:M0})), without corrections is circled by a dotted line. Crossed lines are on mass shell. Here $\Delta_{1\perp}=\Delta_1-q-q^{\prime}$, $\Delta_{2\perp}=\Delta_2+q+q^{\prime}$, $\hat{t}=k^2=(\Delta_{1\perp}-q_1-p_3+q_2)^2$, $\hat{u}=(\Delta_{1\perp}-q_1-p_4)^2$, $\hat{s}=(p_3+p_4-q_1-q_2)^2$.}
	\end{center}
\end{figure}

 The $2\rightarrow 4$ process $p(p_a)+p(p_b)\to p(p_1)+\pi(p_3)+\pi(p_4)+p(p_2)$ can be
described as follows (the notation for any momentum is $k=(k_0,k_z;\vec{k})$, $\vec{k}=(k_x,k_y)$):
\begin{eqnarray}
 p_a&=&\left( 
\frac{\sqrt{s}}{2}, \beta\frac{\sqrt{s}}{2}; \vec{0} 
\right),\;
p_b=\left( 
\frac{\sqrt{s}}{2}, -\beta\frac{\sqrt{s}}{2}; \vec{0} 
\right),\nonumber\\
p_{1,2}&=&
\left( 
E_{1,2},p_{1,2z}; \vec{p}_{1,2\perp}
\right),
E_{1,2}=\sqrt{p_{1,2z}^2+\vec{p}_{1,2\perp}^2+m_p^2},\;\nonumber\\
p_{3,4}&=&
\left( 
m_{3,4\perp}\mathrm{ch}\;y_{3,4}, m_{3,4\perp}\mathrm{sh}\;y_{3,4}; \vec{p}_{3,4\perp}
\right) = \nonumber\\
&=&\left( 
\sqrt{m_{\pi}^2+\vec{p}_{3,4\perp}^2\mathrm{ch}^2\eta_{3,4}}, |\vec{p}_{3,4\perp}|\;\mathrm{sh}\;\eta_{3,4}; \vec{p}_{3,4\perp}
\right), \nonumber\\
m_{i\perp}^2&=&m_i^2+\vec{p}_{i\perp}^2,\; m_{1,2}=m_p,\; m_{3,4}=m_{\pi}, \nonumber\\ 
\vec{p}_{4\perp}&=&-\vec{p}_{3\perp}-\vec{p}_{1\perp}-\vec{p}_{2\perp},\;
\nonumber\\
\beta&=&\sqrt{1-\frac{4m_p^2}{s}},\; s=(p_a+p_b)^2,\; s^{\prime}=(p_1+p_2)^2.
\label{eq2:kinmomenta}
\end{eqnarray}
Here the $y_i$ ($\eta_i$) are the rapidities (pseudorapidities) of the final pions.

 The phase space of the process in terms of the above variables is the following
\begin{eqnarray}
 \mathrm{d}\mathrm{\Phi}_{2\to 4} &=& 
 \left(2\pi\right)^4\delta^4\left( p_a+p_b-\sum_{i=1}^4p_i\right) \prod_{i=1}^4 \frac{\mathrm{d}^3p_i}{(2\pi)^32E_i}=
 \nonumber\\
 &=& \frac{1}{2^4(2\pi)^8}
 \prod_{i=1}^3 p_{i\perp}\mathrm{d}p_{i\perp}\mathrm{d}\phi_i\cdot \mathrm{d}y_3\mathrm{d}y_4 \cdot {\cal J};
  \nonumber\\
 {\cal J}&=& \frac{\mathrm{d}p_{1z}}{E_1} \frac{\mathrm{d}p_{2z}}{E_2} 
 \delta\left( \sqrt{s}-\sum_{i=1}^4 E_i\right)
 \delta\left( \sum_{i=1}^4 p_{iz}\right)=\nonumber\\
 &=&\frac{1}{\left| \tilde{E}_2\tilde{p}_{1z}-\tilde{E}_1\tilde{p}_{2z}\right|},
\label{eq3:kinPhSp}  
\end{eqnarray} 
where the $p_{i\perp}=\left| \vec{p}_{i}\right|$, $\tilde{p}_{1,2z}$ are the appropriate roots of the system
\begin{equation}
\begin{cases}
\label{eq4:sys} &A=\sqrt{s}-E_3-E_4=\sqrt{m_{1\perp}^2+p_{1z}^2}+\sqrt{m_{2\perp}^2+p_{2z}^2},\\
 &B=-p_{3z}-p_{4z}=p_{1z}+p_{2z},
\end{cases}
\end{equation}
\begin{eqnarray}
 \tilde{p}_{1z}&=& \frac{B}{2}+\frac{1}{2(A^2-B^2)}\left[ 
B\left( m_{1\perp}^2-m_{2\perp}^2\right) + A\cdot\lambda_0^{1/2}
\right],\nonumber\\
\label{eq4:sysroot} \lambda_0&=&\lambda\left( A^2-B^2,m_{1\perp}^2,m_{2\perp}^2\right). 
\end{eqnarray}
Here $\lambda(x,y,z)=x^2+y^2+z^2-2xy-2xz-2yz$, and then ${\cal J}=\lambda_0^{1/2}/2$.

For the differential cross-section we have
\begin{eqnarray}
\frac{\mathrm{d}\sigma_{2\to 4}}{\prod_{i=1}^{3\phantom{I}} \mathrm{d}p_{i\perp}\mathrm{d}\phi_i\cdot \mathrm{d}y_3\mathrm{d}y_4 }&=&
\frac{1}{2\beta s}\cdot\frac{\prod_{i=1}^3 p_{i\perp}}{2^4(2\pi)^8\cdot\frac{1}{2}\lambda_0^{1/2}}
\left| T\right|^2=
\nonumber\\
\label{eq5:dcsdall} &=& \frac{\prod_{i=1}^3 p_{i\perp}}{2^{12}\pi^8\beta s \lambda_0^{1/2}}\left| T \right|^2.
\end{eqnarray}
The pseudorapidity is more convenient experimental variable, and we can use the transform
\begin{equation}
\frac{\mathrm{d}y_i}{\mathrm{d}\eta_i}=\frac{p_{i\perp}\mathrm{ch}\eta_i}{\sqrt{m_i^2+p_{i\perp}^2\mathrm{ch}^2\eta_i}}
\end{equation}
to get the differential cross-section in the pseudorapidities.

For exact calculations of the elastic subprocesses (see Fig.~\ref{fig4:totalkin}) of the type\\ $a(p_1)+b(p_2)\rightarrow c(p_1-q_{el})+d(p_2+q_{el})$:
\begin{eqnarray}
q_{el}&=&\left( q_0,q_z; \vec{q}\right),\nonumber\\
q_z&=&-\frac{b}{2a}\left( 1-\sqrt{1-\frac{4ac}{b^2}}\right),\nonumber\\ q_0&=&\frac{A_0q_z+\vec{p}_{1\perp}\vec{q}+\vec{p}_{2\perp}\vec{q}}{A_z},\nonumber\\
a&=&A_z^2-A_0^2,\; b=-2
\left( 
A_z\cdot {\cal D}+A_0 \left( \vec{p}_{1\perp}\vec{q}+\vec{p}_{2\perp}\vec{q}\right)
\right),\nonumber\\
c&=&2 A_z B_z-\left( \vec{p}_{1\perp}\vec{q}+\vec{p}_{2\perp}\vec{q}\right)^2+\vec{q}^2A_z^2,\nonumber\\
A_0&=&p_{1z}+p_{2z},\; A_z=p_{10}+p_{20}, \nonumber\\
B_0&=&p_{1z}\cdot\vec{p}_{2\perp}\vec{q}-p_{2z}\cdot\vec{p}_{1\perp}\vec{q},\nonumber\\
B_z&=&p_{10}\cdot\vec{p}_{2\perp}\vec{q}-p_{20}\cdot\vec{p}_{1\perp}\vec{q},\nonumber\\
{\cal D}&=&p_{1z}p_{20}-p_{2z}p_{10},\label{eq6:qel}
\end{eqnarray}
and $q_{el}^2\simeq -\vec{q}^2$.

\section*{Appendix B. Regge-eikonal model for elastic pro\-ton--pro\-ton and pion--pro\-ton scattering}
\label{app:godizovmodel}

Here is a short review of the formulas for the Regge-eikonal approach~\cite{godizovpp}, \cite{godizovpip}, which we use to estimate rescattering corrections in the proton--proton and pion--proton channels.

The amplitudes of elastic proton--proton and pion--proton scattering are expressed in terms of the eikonal functions:
\begin{eqnarray}
T_{pp,\pi p}^{el}(s,b)&=&\frac{\mathrm{e}^{-2\Omega_{pp,\pi p}^{el}(s,b)}-1}{2\mathrm{i}},\nonumber\\
\Omega_{pp, \pi p}^{el}(s,b) &=& -\mathrm{i}\,\delta_{pp,\pi p}^{el}(s,b),\nonumber\\
\delta^{el}_{pp,\pi p}(s,b)&=&\frac{1}{16\pi s}\int_0^\infty d(-t) J_0(b\sqrt{-t}) \delta^{el}_{pp,\pi p}(s,t),
\label{eq:elamplitudes}
\end{eqnarray}


\begin{eqnarray}
&& \delta^{el}_{pp}(s,t)\simeq\nonumber\\
&& g_{pp\mathbb{P}}(t)^2\left(
\mathrm{i}+\tan\frac{\pi(\alpha_{\mathbb{P}}(t)-1)}{2}) 
\right) \pi \alpha^{\prime}_{\mathbb{P}}(t)
\left( \frac{s}{2s_0}\right)^{\alpha_{\mathbb{P}}(t)},\nonumber\\
&& \alpha_{\mathbb{P}}(t)=1+\frac{\alpha_{\mathbb{P}}(0)-1}{1-\frac{t}{\tau_a}}\,,\, g_{pp\mathbb{P}}(t)=\frac{g_{pp\mathbb{P}}(0)}{\left( 1-a_g t\right)^2}.
\label{eq:godizovpp}
\end{eqnarray}

\begin{eqnarray}
&& \delta^{el}_{\pi p}(s,t)\simeq\nonumber\\
&& \left(
\mathrm{i}+\tan\frac{\pi(\alpha_{\mathbb{P}}(t)-1)}{2}) 
\right) \beta_{\mathbb{P}}(t)
\left( \frac{s}{s_0}\right)^{\alpha_{\mathbb{P}}(t)},\nonumber\\
&& +\left(
\mathrm{i}+\tan\frac{\pi(\alpha_{f}(t)-1)}{2}) 
\right) \beta_{f}(t)
\left( \frac{s}{s_0}\right)^{\alpha_{f}(t)},
\label{eq:godizovpip1}
\end{eqnarray}

\begin{eqnarray}
&& \alpha_{\mathbb{P}}(t)=1+p_1\left[ 
1-p_2 t\left( 
\arctan\left( p_3-p_2t\right) 
-\frac{\pi}{2}
\right)
\right],\nonumber\\
&& \alpha_f(t)=\left(
\frac{8}{3\pi}\gamma(\sqrt{-t+c_f})
\right)^{1/2},\nonumber\\
&& \gamma(\mu)=\frac{4\pi}{11-\frac{2}{3}n_f}\left( 
\frac{1}{\ln\frac{\mu^2}{\Lambda^2}}+\frac{1}{1-\frac{\mu^2}{\Lambda^2}}
\right),\nonumber\\
&& \beta_{\mathbb{P}}(t)=B_{\mathbb{P}}\mathrm{e}^{b_{\mathbb{P}}t}(1+d_1t+d_2t^2+d_3 t^3+d_4 t^4),\nonumber\\
&& \beta_f(t)=B_f \mathrm{e}^{b_f t}.
\label{eq:godizovpip2}
\end{eqnarray}
The parameters can be found in Tables~\ref{tab1} and~\ref{tab2}.

\begin{table}[ht!]
	\caption{\label{tab1} Parameters for proton--proton elastic scattering amplitude.}
	\begin{center}
		\begin{tabular}{|c|c|}
			\hline
			Parameter &   Value  \\	
			\hline	
			$\alpha_{\mathbb{P}}(0)-1$ & $0.109$  \\	           	
			\hline
	$\tau_a$ &  $0.535$~GeV$^2$  \\	           	
	\hline	
	$g_{pp\mathbb{P}}(0)$ & $13.8$~GeV  \\	           	
	\hline				
	$a_g$ &    $ 0.23$~GeV$^{-2}$  \\	           	
	\hline					
		\end{tabular}
	\end{center}
\end{table}  

\begin{table}[ht!]
	\caption{\label{tab2} Parameters for pion--proton elastic scattering amplitude.}
	\begin{center}
		\begin{tabular}{|c|c|}
			\hline
			Parameter &   Value  \\	
			\hline	
			$B_{\mathbb{P}}$ & $26.7$  \\	           	
			\hline
			$b_{\mathbb{P}}$ &  $2.36$~GeV$^{-2}$  \\	           	
			\hline	
			$d_1$ & $0.38$~GeV$^{-2}$  \\	           	
			\hline				
			$d_2$ &    $0.3$~GeV$^{-4}$  \\	           	
			\hline	
			$d_3$ &    $-0.078$~GeV$^{-6}$  \\	           	
			\hline
			$d_4$ &    $0.04$~GeV$^{-8}$  \\	           	
			\hline
			$B_f$ &    $67$  \\	           	
			\hline
			$b_f$ &    $1.88$~GeV$^{-2}$  \\	           	
			\hline														
		\end{tabular}
	\end{center}
\end{table} 

We have
\begin{eqnarray}
V_{pp}(s,q^2)&=&\int d^2\vec{b} \,\mathrm{e}^{\mathrm{i}\vec{q}\vec{b}}
\sqrt{1+2\mathrm{i}T_{pp}^{el}(s,b)}=\nonumber\\
&=& \int d^2\vec{b}\, \mathrm{e}^{\mathrm{i}\vec{q}\vec{b}} 
\mathrm{e}^{-\Omega_{pp}^{el}(s,b)}=\nonumber\\
&=& (2\pi)^2\delta^2\left( \vec{q}\right) + 2\pi \bar{T}_{pp}(s,q^2),
\label{eq:Vpp1}\\
\bar{T}_{pp}(s,q^2)&=& \int_0^\infty b\,db\, J_0\left( b\sqrt{-q^2}\right)
\left[ 
\mathrm{e}^{-\Omega_{pp}^{el}(s,b)}-1
\right]
\label{eq:Vpp2}\\
S_{\pi p}(s,q^2)&=&\int d^2\vec{b} \,\mathrm{e}^{\mathrm{i}\vec{q}\vec{b}}
\left(
1+2\mathrm{i}T_{\pi p}^{el}(s,b)
\right)=\nonumber\\
&=& \int d^2\vec{b}\, \mathrm{e}^{\mathrm{i}\vec{q}\vec{b}} 
\mathrm{e}^{-2\Omega_{\pi p}^{el}(s,b)}=\nonumber\\
&=& (2\pi)^2\delta^2\left( \vec{q}\right) + 2\pi \bar{T}_{\pi p}(s,q^2),
\label{eq:Vpip1}\\
\bar{T}_{\pi p}(s,q^2)&=& \int_0^\infty b\,db\, J_0\left( b\sqrt{-q^2}\right)
\left[ 
\mathrm{e}^{-2\Omega_{\pi p}^{el}(s,b)}-1
\right]
\label{eq:Vpip2}
\end{eqnarray}

Here we take $S_{\pi p}(s,t)=S_{\pi^+ p}(s,t)=S_{\pi^- p}(s,t)$ and
\begin{equation}
T^{el}_{\pi^+p}(s,t)=T^{el}_{\pi^-p}(s,t)=4\pi s \bar{T}_{\pi p}(s,t)
\label{eq:Tpipexact}
\end{equation}

The approach~(\ref{eq:godizovpip1}) describes the data on pion--proton scattering better even at low energies, that
is why we use it instead of the one presented in~\cite{godizovpp}.

The functions $\bar{T}_{pp}$ and $\bar{T}_{\pi p}$ are convenient for numerical
calculations, since the oscillations are not so strong.

\section*{Appendix C. Covariant basis and Po\-me\-ron--Po\-me\-ron to di-pion cross-section}
\label{app:covreg}

In the classical covariant reggeization, as was considered, for example, in~\cite{oldcovreg}, and in the author's  papers~\cite{mySD},\cite{myCEDP2}, we
have the following structure of the amplitudes. Basic elements of 
such an approach are the vertex functions \linebreak
$T^{\mu_1\ldots\mu_{J}}(p,q)$, where
\begin{equation}
 \label{eq:TvertexDef}
 T^{\mu_1\dots\mu_J}(p,q)=<p-q| I^{\mu_1\dots\mu_J}|p>,
\end{equation}
hadronic tensor
\begin{eqnarray}
&& W^{\mu_1\ldots\mu_J\nu_1\ldots\nu_{J^{\prime}}}(p,q)=\nonumber\\
&&\int d^4x\; e^{iqx}
\left< p|\; I^{\mu_1\ldots\mu_J}(x)I^{\nu_1\ldots\nu_{J^{\prime}}}(0)\; |p \right>\quad{,}
\label{eq:WtensorDef}
\end{eqnarray}
and propagators $\Pi_{\mu_1\ldots\mu_J,\;\nu_1\ldots\nu_{J}}(J,t)/(m^2(J)-t)$
with the tensor structure $\Pi_{\mu_1\ldots\nu_J}$ calculated in~\cite{oldcovreg}, for 
example. $1/(m^2(J)-t)$ have the poles at
\begin{equation}
\label{eq:polesJ}
m^2(J)-t=0,\; \mbox{i.e.}\; J=\alpha_{{\mathbb P}}(t)\;,
\end{equation}
after an appropriate analytic continuation of the signatured amplitudes 
in $J$. We assume that this pole, where $\alpha_{\mathbb P}$ is the Pomeron 
trajectory, gives, by definition,  the dominant contribution at high energies 
after having taken the corresponding residues. At this stage
we do not take into account absorptive corrections (unitarization).

$I^{\mu_1\dots\mu_{J}}$ is the current operator related to the hadronic spin-$J$ Heisenberg field operator,
\begin{equation}
\label{eq:KGJ}
\left( \square + m_J^2\right) \Phi^{\mu_1\dots\mu_J}(x)=I^{\mu_1\dots\mu_J}(x)\;,
\end{equation}
and
\begin{eqnarray}
&&\partial_{\mu}I^{\mu_1\ldots\mu\ldots\mu_J}=
\partial_{\nu}I^{\nu_1\ldots\nu\ldots\nu_{J^{\prime}}}=0\;{;}\label{eq:Icond1}\\
&&I^{\mu_1\ldots\mu_J}=I^{\left(\mu_1\ldots\mu_J\right)}\;{;}\;
I^{\nu_1\ldots\nu_{J^{\prime}}}=I^{\left(\nu_1\ldots\nu_{J^{\prime}}\right)}\;{;}
\label{eq:Icond2}\\
&&g_{\mu_i\mu_k}I^{\mu_1\ldots\mu_i\ldots\mu_k\ldots\mu_J}=
g_{\nu_i\nu_k}I^{\nu_1\ldots\nu_i\ldots\nu_k\ldots\nu_{J^{\prime}}}=0.\label{eq:Icond3}
\end{eqnarray}
In momentum space the Rarita-Schwinger conditions (\ref{eq:Icond1})--(\ref{eq:Icond3}) for the vertex are
\begin{eqnarray}
&&T^{\mu_1\dots\mu_i\dots\mu_j\dots\mu_J}=T^{\mu_1\dots\mu_j\dots\mu_i\dots\mu_J}\label{eq:Tcond1}\\
&&q_{\mu_i}T^{\mu_1\dots\mu_i\dots\mu_J}=0\label{eq:Tcond2}\\
&&g_{\mu_i\mu_j}T^{\mu_1\dots\mu_i\dots\mu_j\dots\mu_J}=0\label{eq:Tcond3},
\end{eqnarray}
and the same conditions are imposed on each group of indices in the tensors $W$ and $\Pi$. Let us 
note (as was done in~\cite{oldcovreg}), that conditions~(\ref{eq:Tcond1})-(\ref{eq:Tcond3}) are
valid only on the mass shell of the spin-J meson. And when we go to the phase space of the scattering region, these
conditions could be relevant only for conserved hadronic currents. However, this may well not be the case.

Let us consider both cases. In the case of conserved currents we can define the main transverse structures:
\begin{eqnarray}
 G_{\alpha\beta}&=&g_{\alpha\beta}-\frac{q_{\alpha}q_{\beta}}{q^2}\;{;}\nonumber\\
 P_{\alpha}&=&\left( p_{\alpha}-q_{\alpha}\frac{pq}{q^2}\right)/
\sqrt{m^2-(pq)^2/q^2},\;P^2=1,\nonumber\\
 K_{\alpha}&=&\left( k_{\alpha}-q_{\alpha}\frac{kq}{q^2}\right)/
\sqrt{m^2-(kq)^2/q^2},\;K^2=1;\nonumber\\
 G_{\alpha\beta}P^{\beta}&=&P_{\alpha},\; G_{\alpha\beta}K^{\beta}=K_{\alpha},\nonumber\\
 g_{\alpha\beta}G^{\alpha\beta}&=&G_{\alpha\beta}G^{\alpha\beta}=3.
\label{eq:structures}
\end{eqnarray} 
For the vertex
functions $T$ we can obtain the following tensor decomposition:
\begin{eqnarray}
&& T^{(J)}\equiv T^{\mu_1\dots\mu_{J_i}}(k,q)=\nonumber\\
&&\phantom{T^{(J)}\equiv}\label{eq:Ttensor0} F_J(t) 
\sum_{n=0}^{\left[\frac{J}{2}\right]} 
\frac{{\mathbb C}_J^n}{{\mathbb C}_J^0} \left( K^{(J-2n)}G^{(n)}\right)\;,\\
&&\phantom{T^{(J)}\equiv}\label{eq:CJn0}{\mathbb C}_J^n=
\frac{(-1)^n (2(J-n))\mbox{!}}{(J-n)\mbox{!}n\mbox{!}(J-2n)\mbox{!}},
\end{eqnarray}
where the 
tensor structures $\left( K^{(J-2n)}G^{(n)}\right)^{\mu_1\dots\mu_{J}}$ satisfy only the two 
conditions~(\ref{eq:Tcond1}) and (\ref{eq:Tcond2}) (transverse-symmetric):
\begin{eqnarray}
&&\left( K^{(J-2n)}G^{(n)}\right)=\nonumber\\
&&\label{KGsym} \frac{K^{\; (\mu_1}\!\!\cdot_{\cdots}\!\!\cdot K^{\; \mu_{J-2n}}
G^{\mu_{J-2n+1}\mu_{J-2n+2}}\!\!\cdot_{\cdots}\!\!\cdot G^{\mu_{J-1}\mu_{J})}}{N_{J}^n},\\
&& \label{coefPGsym}N_{J}^n=\frac{J\mbox{!}}{2^n n\mbox{!}(J-2n)\mbox{!}}.
\end{eqnarray}
The coefficients ${\mathbb C}_{J}^n$ in~(\ref{eq:Ttensor0}) can be obtained from 
the condition~(\ref{eq:Tcond3}) which
leads to the recurrent set of equations (see~\cite{myCEDP2}). It was also shown in~\cite{myCEDP2} that,
for elastic scattering of particles with equal masses, which can be obtained by the contraction 
$T_{\{\mu\}}(p_a,\Delta)\otimes T_{\{\mu\}}(p_b,-\Delta)$, we have the usual Regge expression for the amplitude. In the general elastic process
$a+b\to c+d$ with unequal masses of particles we can obtain
\begin{eqnarray}
&&\!\!\!\!\!{\cal M}(s,t)=T^{(J)}_{\{\mu\}}(p_a,\Delta)\otimes 
T^{(J)}_{\{\mu\}}(p_b,-\Delta)=\nonumber\\
&&\!\!\!\!\! = F^{(1)}_J(t)F^{(2)}_J(t) 2^J \times\nonumber\\
&&\!\!\!\!\! 
{\cal P}_J\left( 
\frac{(s-m_a^2-m_b^2+\frac{(m_a^2-m_c^2+t)(m_b^2-m_d^2+t)}{2t})(-2t)}{\lambda^{1/2}(m_b^2,m_d^2,t)\;\lambda^{1/2}(m_a^2,m_c^2,t)}
\right)\label{UneqMass}
\end{eqnarray}
Here the argument of the Legendre function is the t-channel cosine $z_t=\cos\theta_t$, and 
$$
\lambda(x,y,z)=x^2+y^2+z^2-2xy-2xz-2yz.
$$
In the classical Regge scheme
\begin{equation}
\sum_J (2J+1){\cal M}_J {\cal P}_J(-z_t)\to
\eta_{\mathbb R}(t)
\beta_{\mathbb R}(t) \left( \frac{s}{s_0}\right)^{\alpha_{\mathbb R}(t)},
\label{eq:ReggeClassic}
\end{equation}
where we also have Legendre polynomials.

In the case of non-conserved currents we have no Rarita--Schwinger conditions and could propose
only some arguments on behavior of coefficients in tensors
\begin{eqnarray}
&&\label{eq:Tnonconserv} 
T^{(J)}_{\mu_1\ldots\mu_J}(p,q)=
\sum_{n+k\le J}f_J^{n,k} \left\{ p^n q^k g^{\left[ (J-n-k)/2\right]} \right\},
\\
&&\label{eq:Propnonconserv}
\Pi_{\mu_1\ldots\mu_J,\;\nu_1\ldots\nu_J}=Q_{(\mu_1\nu_1}\ldots Q_{\mu_J\nu_J)}+\mathrm{nonleading} ,
\end{eqnarray}
where 
\begin{eqnarray}
&&p^n=p_{\mu_1}\ldots p_{\mu_n}, q^k=q_{\mu_{n+1}}\ldots p_{\mu_{n+k}},\nonumber \\
&&g^{\left[ (J-n-k)/2\right]}=g_{\mu_{n+k+1}\mu_{n+k+2}}\ldots g_{\mu_{J-1}\mu_J},\nonumber
\end{eqnarray}
the tensors $\left\{p^n q^k g^{\left[ (J-n-k)/2\right]}\right\}$ and $Q_{(\mu_1(\nu_1}\ldots Q_{\mu_J)\nu_J)}$ are 
symmetric on the $\mu_i$ and $\nu_j$ groups of indices. These arguments are
\begin{enumerate}
\item all $f_J^{n,k}$ are of the same order of magnitude;
\item the tensor $\Pi$ has the form~(\ref{eq:Propnonconserv}), where 
$$
Q_{\mu\nu}=g_{\mu\nu}-q_{\mu}q_{\nu}/m_J^2,
$$ 
which is equal to $G_{\mu\nu}$ on the mass shell of the spin-J meson, and other terms in~(\ref{eq:Propnonconserv}) give
nonleading contributions to the final result.
\end{enumerate}
In this case we have also a simple Regge result like in~(\ref{eq:ReggeClassic}), but without Legendre functions. It is natural to assume that, when the spin-J meson is not far from the mass shell, the structures of vertex and propagator are close
to the case of conserved currents. 


\end{document}